\documentclass[%
reprint,
superscriptaddress,
 amsmath,amssymb,
 aps,
]{revtex4-2}

\usepackage{graphicx}% Include figure files
\usepackage{dcolumn}% Align table columns on decimal point
\usepackage{bm}% bold math
\usepackage[T1]{fontenc}
\usepackage[utf8]{inputenc}
\usepackage{textcomp}
\begin{document}

\preprint{APS/123-QED}

\title{Controllable Spin-Resolved Photon Emission Enhanced by Slow-Light Mode in Photonic Crystal Waveguides on Chip}% Force line breaks with \\

\author{Shushu Shi}
\author{Shan Xiao}
\affiliation{Beijing National Laboratory for Condensed Matter Physics, Institute of Physics, Chinese Academy of Sciences, Beijing 100190, China}
\affiliation{CAS Center for Excellence in Topological Quantum Computation and School of Physical Sciences, University of Chinese Academy of Sciences, Beijing 100049, China}
\author{Jingnan Yang}
\affiliation{State Key Laboratory for Mesoscopic Physics and Frontiers Science Center for Nano-optoelectronics, School of Physics, Peking University, 100871 Beijing, China}
\author{Shulun Li}
\affiliation{State Key Laboratory for Superlattice and Microstructures, Institute of Semiconductors,Chinese Academy of Sciences, Beijing, 100083, China}
\author{Xin Xie}
\author{Jianchen Dang}
\author{Longlong Yang}
\author{Danjie Dai}
\affiliation{Beijing National Laboratory for Condensed Matter Physics, Institute of Physics, Chinese Academy of Sciences, Beijing 100190, China}
\affiliation{CAS Center for Excellence in Topological Quantum Computation and School of Physical Sciences, University of Chinese Academy of Sciences, Beijing 100049, China}
\author{Bowen Fu}
\affiliation{State Key Laboratory for Mesoscopic Physics and Frontiers Science Center for Nano-optoelectronics, School of Physics, Peking University, 100871 Beijing, China}
\author{Sai Yan}
\author{Yu Yuan}
\author{Rui Zhu}
\affiliation{Beijing National Laboratory for Condensed Matter Physics, Institute of Physics, Chinese Academy of Sciences, Beijing 100190, China}
\affiliation{CAS Center for Excellence in Topological Quantum Computation and School of Physical Sciences, University of Chinese Academy of Sciences, Beijing 100049, China}
\author{Bei-Bei Li}
\affiliation{Beijing National Laboratory for Condensed Matter Physics, Institute of Physics, Chinese Academy of Sciences, Beijing 100190, China}
\affiliation{Songshan Lake Materials Laboratory, Dongguan, Guangdong 523808, China}
\author{Zhanchun Zuo}
%\email{zczuo@iphy.ac.cn}
\affiliation{Beijing National Laboratory for Condensed Matter Physics, Institute of Physics, Chinese Academy of Sciences, Beijing 100190, China}
\affiliation{CAS Center for Excellence in Topological Quantum Computation and School of Physical Sciences, University of Chinese Academy of Sciences, Beijing 100049, China}
\author{Can Wang}
\email{canwang@iphy.ac.cn}
\affiliation{Beijing National Laboratory for Condensed Matter Physics, Institute of Physics, Chinese Academy of Sciences, Beijing 100190, China}
\affiliation{CAS Center for Excellence in Topological Quantum Computation and School of Physical Sciences, University of Chinese Academy of Sciences, Beijing 100049, China}
\affiliation{Songshan Lake Materials Laboratory, Dongguan, Guangdong 523808, China}
\author{Haiqiao Ni}
\author{Zhichuan Niu }
\affiliation{State Key Laboratory for Superlattice and Microstructures, Institute of Semiconductors,Chinese Academy of Sciences, Beijing, 100083, China}
\author{Kuijuan Jin}
\affiliation{Beijing National Laboratory for Condensed Matter Physics, Institute of Physics, Chinese Academy of Sciences, Beijing 100190, China}
\affiliation{CAS Center for Excellence in Topological Quantum Computation and School of Physical Sciences, University of Chinese Academy of Sciences, Beijing 100049, China}
\affiliation{Songshan Lake Materials Laboratory, Dongguan, Guangdong 523808, China}

\author{Qihuang Gong}
\author{Xiulai Xu}
\email{xlxu@pku.edu.cn}
\affiliation{State Key Laboratory for Mesoscopic Physics and Frontiers Science Center for Nano-optoelectronics, School of Physics, Peking University, 100871 Beijing, China}
\date{\today}% It is always \today, today,
             %  but any date may be explicitly specified
\begin{abstract}
We report the slow-light enhanced spin-resolved in-plane emission from a single quantum dot (QD) in a photonic crystal waveguide (PCW). The slow light dispersions in PCWs are designed to match the emission wavelengths of single QDs. The resonance between two spin states emitted from a single QD and a slow light mode of a waveguide is investigated under a magnetic field with Faraday configuration. Two spin states of a single QD experience different degrees of enhancement as their emission wavelengths are shifted by combining diamagnetic and Zeeman effects with an optical excitation power control. A circular polarization degree up to 0.81 is achieved by changing the off-resonant excitation power. Strongly polarized photon emission enhanced by a slow light mode shows great potential to attain controllable spin-resolved photon sources for integrated optical quantum networks on chip.
\end{abstract}

\maketitle

\section{\label{sec:level1}Introduction}
Cavity quantum electrodynamics, which investigates light-matter interaction at the single-photon level, holds a lot of potential to realize scalable integrated optical quantum networks and to boost applications in quantum information processing and computing \cite{lodahl2015interfacing,dietrich2016gaas}. In particular, efficient single-photon sources are highly desired as one of the fundamental components of quantum technologies \cite{senellart2017high,arakawa2020progress}. The
enhanced spontaneous emission of quantum emitters has been demonstrated to implement single photon sources with many devices, such as micropillars \cite{santori2002indistinguishable,pelton2002efficient,gazzano2013bright,schlehahn2016electrically}, microdisks \cite{kiraz2001cavity,yang2021enhanced}, microring resonators \cite{dusanowski2020purcell}, nanobeams \cite{kirvsanske2017indistinguishable,hepp2020purcell,gong2010nanobeam}, and planar photonic crystals \cite{englund2005controlling,chang2006efficient,balet2007enhanced,xie2020cavity}. Among these microscopic structures, PCWs are notable for on-chip straightforward channels of single photons \cite{yao2010chip}. The in-plane photon emission in PCWs has higher routing efficiency with less reduction than that in photonic crystal cavities, which usually requires to be subsequently coupled out to waveguides for the transmission \cite{schwagmann2012plane}.

Efficient photon routing values the overall probability of channeling an emitted photon into the propagation mode of waveguides, which can be described by the ${\beta}$ factor \cite{thyrrestrup2010extraction}. The ${\beta}$ factor of a PCW is experimentally demonstrated to maintain almost above 0.9 despite the influence of the position variation of quantum emitters in the slow-light regime \cite{arcari2014near}. Besides, PCWs have relatively large mode volumes and a broad spectral range of dense PCW modes, which enable broadband resonance with quantum emitters \cite{vlasov2005active,laucht2012broadband,lund2008experimental} compared to the Fabry-Pérot (FP) resonances. These advantages consequently loosen the spectral and spatial match requirements for the coupling between quantum emitters and PCWs.

The coupling of a single QD with a PCW has been intensively investigated to achieve enhanced spontaneous single photon emission \cite{arcari2014near,dewhurst2010slow,ba2012enhanced,schwagmann2011chip,laucht2012waveguide,madsen2014efficient}. The efficient enhancement demands precise spectral overlap of a QD and a waveguide mode. The QD emission wavelength can be manipulated by changing the temperature \cite{qian2018two,qian2019enhanced}, applying an electric field \cite{kirvsanske2017indistinguishable,hepp2020purcell,wu2020electron} or magnetic field \cite{kim2011magnetic,cao2016observation,peng2017probing,wu2019anisotropies}. Among these methods, tuning the magnetic field enables the control of spin states of excitons. The enhancement of one preferred exciton spin state from single QDs has been performed in a micropillar cavity with a high circular polarization degree of 0.93 \cite{ren2012spin}. Such a controllable spin-polarized photon source has applications in cryptographic optical communications \cite{bhattacharya2010quantum}. However, the out-of-plane emission from micropillars, which requires to be coupled out, limits the tunable spin-polarized photon source to be ultimately integrated on a chip. Instead, PCWs with in-plane emissions can solve the problem.

Here, we demonstrate a spin-resolved photon emission enhancement through the coupling between a waveguide mode and spin states from a QD, utilizing an external magnetic field and optical excitation power control. We offer a fine adjustment on the frequency detuning between the cavity mode and the target single QD by simply tuning excitation laser intersity. PCWs are designed and fabricated to provide different spectral distributions of the slow light regime, assisted with numerical simulations on dispersion curves of PCWs. We tune the spectral overlap between a waveguide mode and two exciton spin states from a single QD to achieve the selective enhancement on spin-resolved emissions. The intensity contrast of spin state emissions with a circular polarization degree up to 0.81 is experimentally demonstrated. This work is beneficial for realizing a controllable in-plane spin-polarized photon source, which has potential applications for on-chip optical quantum networks.
\section{\label{sec:level2}Slow-light waveguide modes}
The sketch of a PCW device with a line defect located in the slab center and two grating couplers terminated on both sides is shown in Fig. 1(a). The QDs are non-resonantly excited as light sources in the center (the green region) of a PCW by a continuous-wave laser with a wavelength of 532 nm. At the same time, the transmission spectra are collected from the grating couplers (the yellow regions). Figure 1(b) depicts the variations of the radii $r_1$ of holes labeled by orange dashed circles in the first row near the waveguide while the radii $r$ of other holes remain unchanged. This design offers a finer adjustment on the dispersion curves of PCWs than that by direct changing the radii $r$ of all holes.
\begin{figure}[t]
	\includegraphics[width=8.6cm]{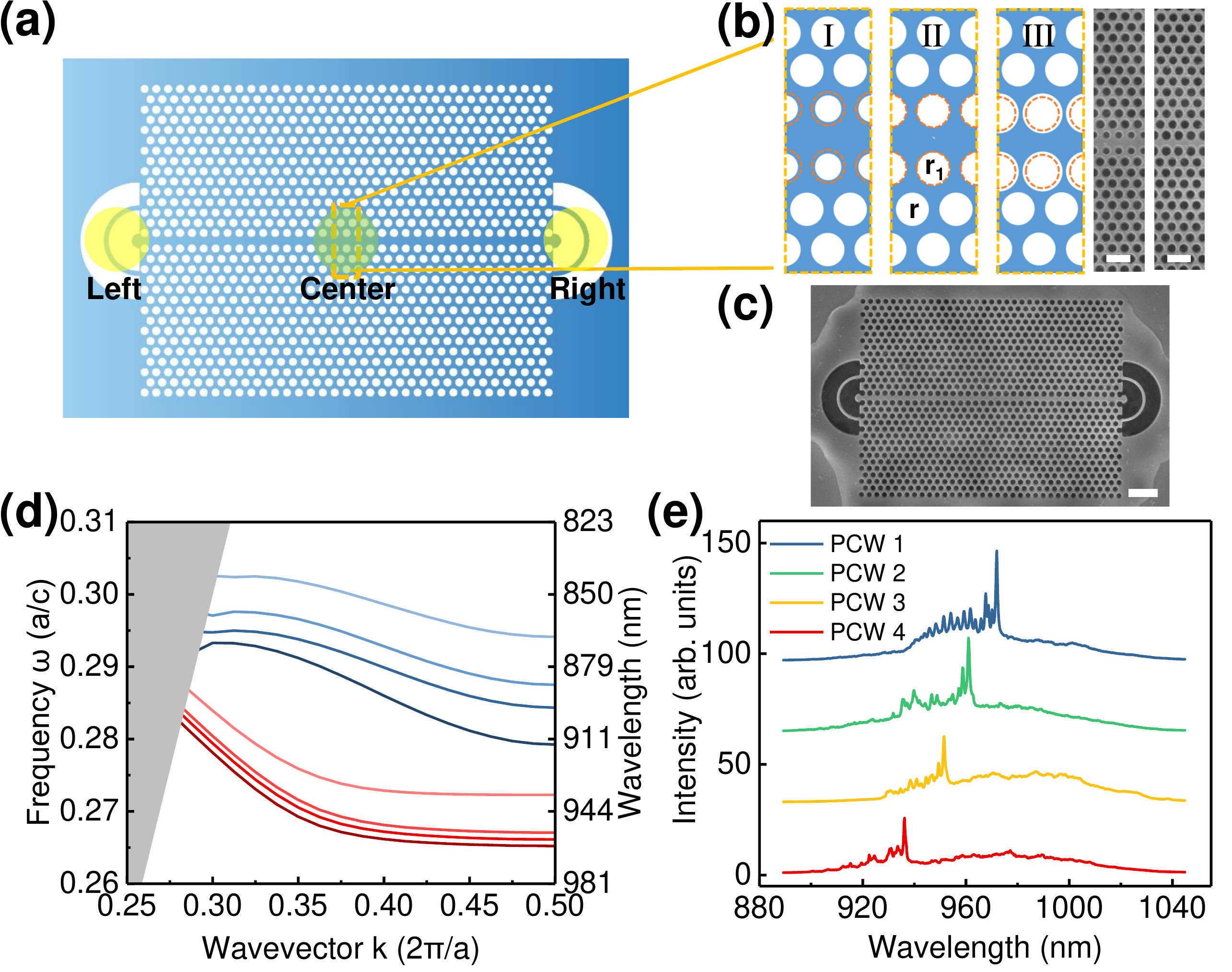}
	\caption{(a) Sketch of the photonic crystal waveguide embedded with quantum dots. (b) Enlarged schematics of designed PCWs. The orange circles indicate the variation of radii ($r_1$) of holes in the first rows. The schematics show three types of waveguides with (I) $r_1<r$, (II) $r_1=r$, (III) $r_1>r$, respectively. The zoom-in SEM images with the scale bar of 0.5 $\mu$m correspond to (I) and (III). (c) SEM image of a PCW. The white scale bar is 1 $\mu$m. The difference between the design sizes of the radii  $r$ and $r_1$ and fabrication results are about 2-5 nm. (d) Dispersion curves of the PCWs with different $r_1$ (increasing by 5 nm for each step) display a frequency change of slow light regime indicated by the fundamental modes. The color depth of the guided modes become lighter as $r_1$ increases. The red lines correspond to the fundamental modes, while the blue lines correspond to the first-order modes. (e) PL spectra of PCW1 to PCW4 with different slow light regime as $r_1$ gradually increased, while $r$ keeps constant at $74$ nm with a relatively high excitation power ($800~{\mu}$w) at low temperature (5 K). The different wavelength distributions of bandwidth for each waveguide match the simulation results in (d). The spectra are shifted for clarity.}
\end{figure}

The Purcell factor of PCWs is inversely proportional to the group velocity of light \cite{hughes2004enhanced}. The group velocity $v_g$ corresponds to the slope of dispersion curves of the fundamental guided mode in PCWs. When the wavelength of light gets closer to the band edge, the group velocity will decrease and cause an increase in the local density of states, leading to a larger Purcell factor and ${\beta}$ factor \cite{rao2007single}. Figure 1(d) presents the simulated dispersion curves of the PCWs possessing the lattice constant $a=255$ nm, $r=0.29a= 74$ nm, $r_1=64$ nm, $69$ nm, $74$ nm, $79$ nm and a refractive index $n=3.4$. The gray area indicates the light cone, and the red curves (the blue curves) correspond to the fundamental mode (the first-order mode) of the guided modes for each waveguide. The frequency of dispersion curves changes when tuning $r_1$. Therefore,  changing $r_1$ in PCWs can tune the waveguide band edge and the slow light regime.

Figure 1(c) shows the scanning electron microscope (SEM) image of the fabricated PCW (PCW4) with a designed waveguide length of $L=9~{\mu}$m, $a=255$ nm, $r=0.29a$ and $r_1=79$ nm. The devices are fabricated with a layer of self-assembled InAs QDs embedded in a 150-nm-thick GaAs slab grown by molecular beam epitaxy with a QD density around $500~{\mu}$m$^{-2}$. The slab is grown on a 1-$\mu$m-thick AlGaAs as a sacrificial layer on a GaAs substrate. The PCW structure is patterned by electron beam lithography and developed to form a pattern mask with exposed holes for inductively coupled plasma etching. Hydrofluoric acid etching is then utilized to remove the sacrificial layer, leaving a suspended membrane of PCWs with different $r_1$, as the zoom-in SEM images in Fig. 1(b) exhibit.

Figure 1(e) shows the photoluminescence (PL) spectra of slow-light waveguide modes from PCW1 to PCW4 with increased $r_1$. The band edges for PCW1-PCW4 range from 972 nm to 936 nm, which correspond well with simulations ranging from 962 nm to 938 nm in Fig. 1(d). The QDs are off-resonance excited by a high-power laser of $800~{\mu}$w at a low temperature of 5 K. The QD ensemble inhomogeneous emission covers a wavelength range from 900 to 1040 nm, which is centered around 960 nm. Therefore, the optical modes of PCW1  show a pronounced feature of slow light in Fig. 1(e), with regular waveguide mode being resolved. PCW4 provides a suitable mode distribution to investigate the coupling with single QDs because the QD density is low within the wavelength regime. It is noted that some modes of PCW4 are not excited due to the transmission bandwidth being out of the wavelength range of QD ensemble emissions. By tuning the wavelength of slow-light modes of PCWs, the effective enhancement can be realized and improved through both weak coupling and slow light effect when a single QD is in resonance with the waveguide modes close to the band edge \cite{arcari2014near}.

\begin{figure}[h]
	\includegraphics[width=7.2cm]{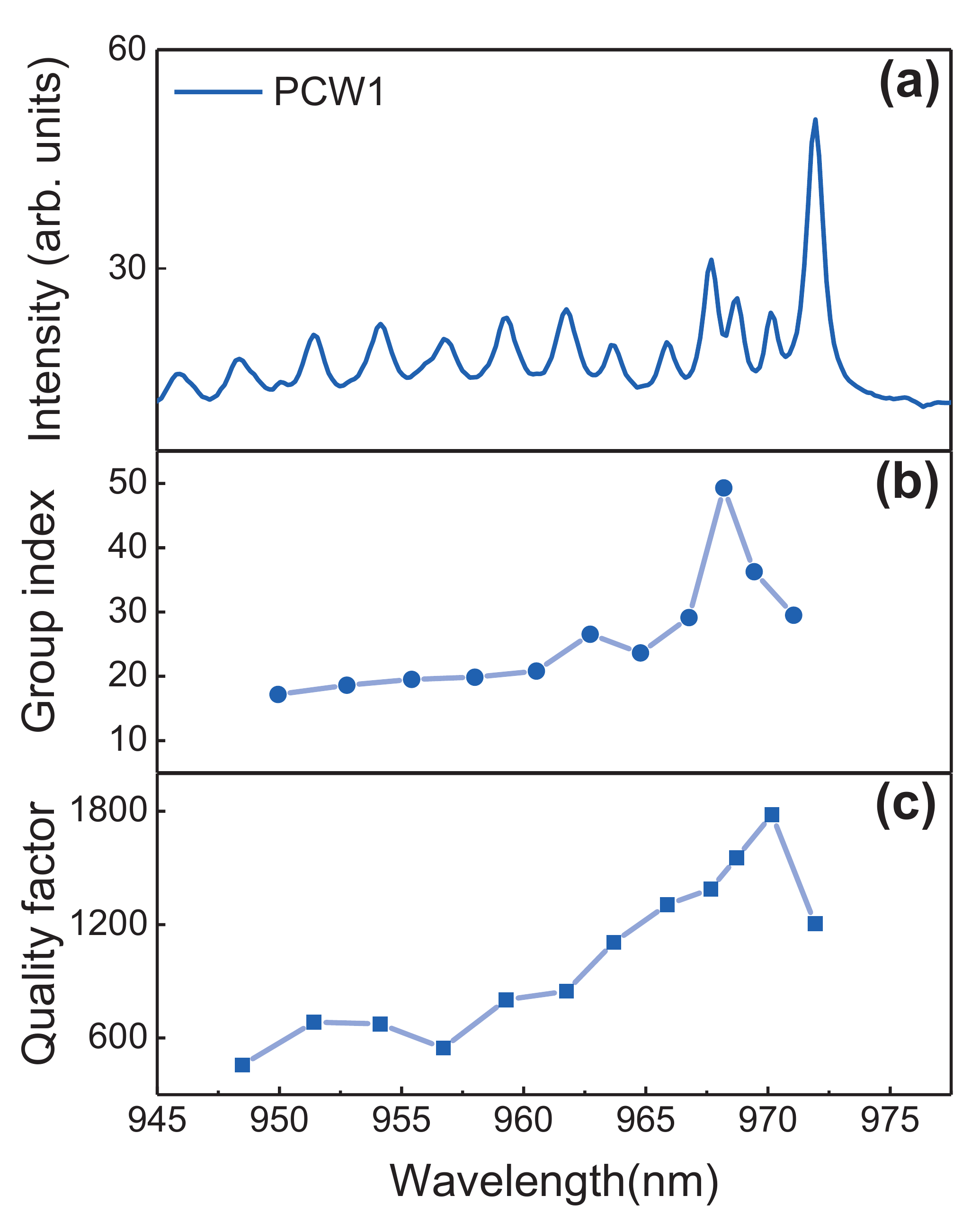}
	\caption{(a) The enlarged PL spectrum of mode distribution from PCW1 in Fig. 1(e). (b) The group index $n_g$ extracted from spectrum in (a) as a function of the emission wavelength. (c) The quality factors for each peak in (a) obtained by Lorentz fitting.}
\end{figure}

We first analyze the mode distribution of PCW1. The enlarged PL spectrum of PCW1 shown in Fig. 2(a) exhibits a transmission bandwidth near 30 nm and a clear decreasing trend of free spectral range (FSR) between adjacent resonant peaks as the emission wavelength increases. Therefore, the group index of PCW1 displays a gradual increase with the increasing wavelength and can be up to 50, as shown in Fig. 2(b), similar to the group index for slow-light waveguides previously reported in Ref. \cite{ek2014slow}. The group index $n_g= c/v_g$, where c is the velocity of light in vacuum, can be quantified by wavelength-dependent variation of resonant peaks as $n_g=\lambda^2/(2{\delta}{\lambda}L)$. The fringe spacing ${\delta}{\lambda}$ and the emission wavelength ${\lambda}$ are obtained by Lorentz fitting of the spectrum in Fig. 2(a), and the length (L) of PCW1 is about 9 $\mu$m. The quality factor is calculated by $Q={\lambda}/{\Delta}{\lambda}$, where ${\Delta}{\lambda}$ is the linewidth of peaks in Fig. 2(a). Figure 2(c) shows that the quality factor increases as the emission wavelength gets close to the band edge of PCW1. The densely distributed slow-light modes provide a higher probability of achieving a spectral match between QDs and a waveguide mode for effective coupling.

\section{\label{sec:level3}Slow light enhanced spin-selected emission}
\begin{figure}[b]
	\includegraphics[width=8.6cm]{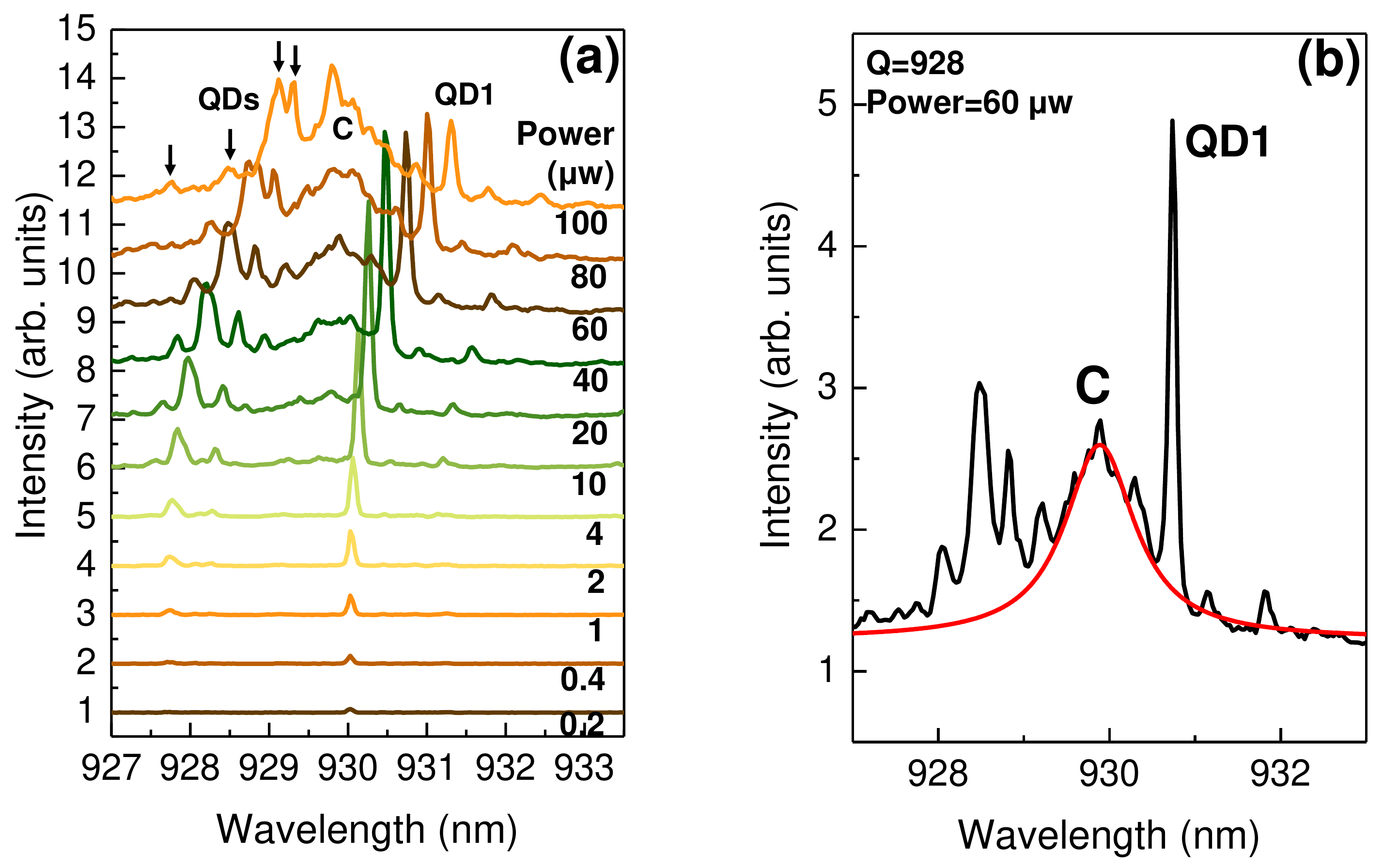}
	\caption{(a) Power-dependent PL spectra of QDs with a waveguide mode C in PCW4. PL peaks from quantum dots redshift with the increasing excitation power, while the waveguide mode redshifts a little by comparison. (b) PL spectrum at 60 $\mu$w from (a) with a fitted Q of about 928 as shown with the red line.}
\end{figure}
\begin{figure*}[htp]
	\includegraphics[width=15cm]{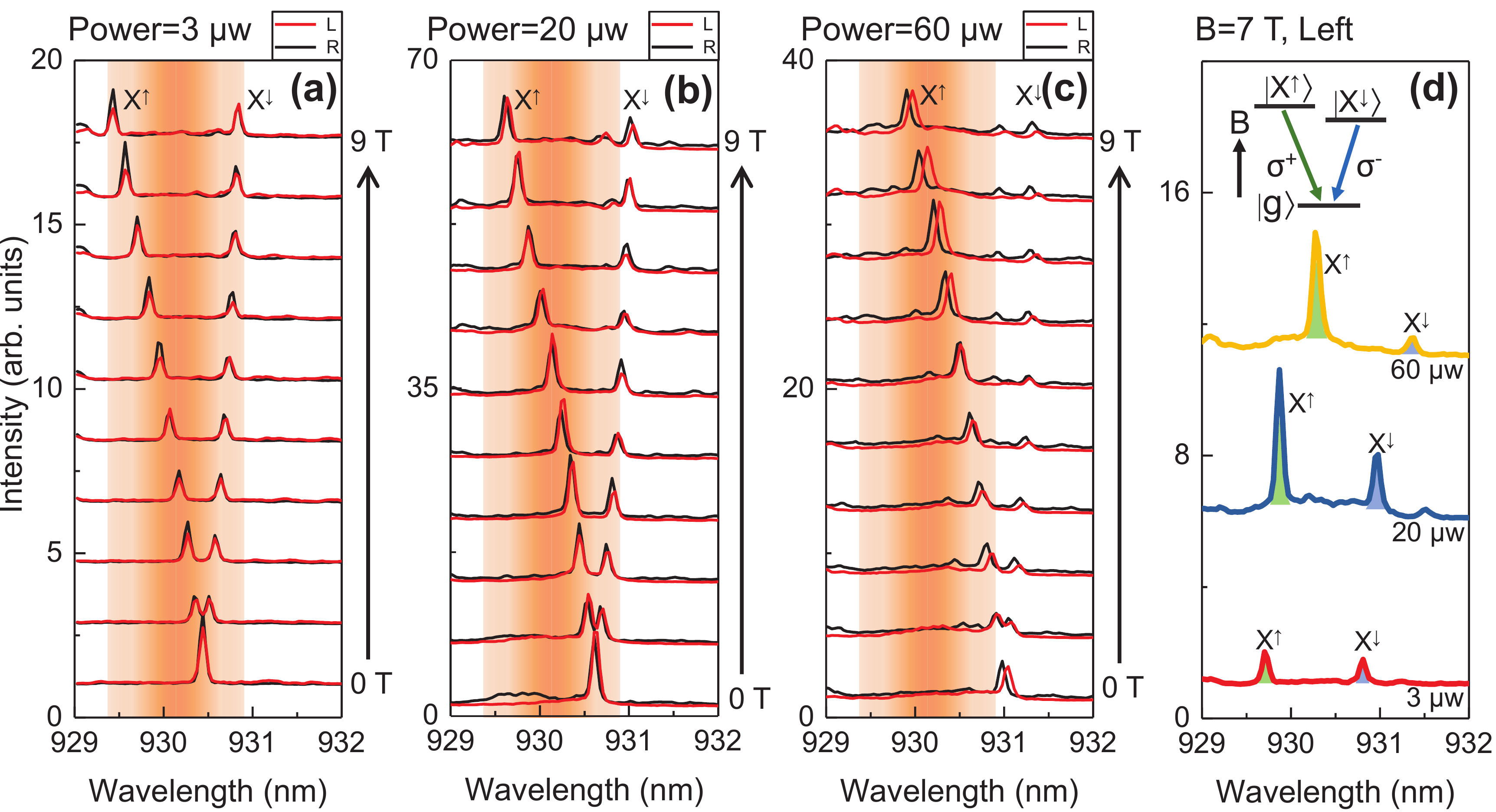}
	\caption{\label{fig:wide}Transmission PL spectra of X$^\uparrow$ and X$^\downarrow$ exciton states of QD1 collected from the left (red lines) and right (black lines) grating coupler with an applied magnetic field in Faraday geometry from 0 to 9 T under different excitation powers of (a) 3 $\mu$w, (b) 20 $\mu$w and (c) 60 $\mu$w. The orange area indicates the position of the waveguide mode C. The PL peaks of each spin slightly redshift due to temperature increase with the increasing excitation power. (d) The PL spectra collected from the left grating coupler of X$^\uparrow$ and X$^\downarrow$ exciton states of QD1 with different excitation powers in a magnetic field of 7 T. Selective enhancement of spin states is clearly demonstrated with different excitation powers. During the PL measurement, the integration time for each spectrum is 2 s for the excitation power of 60 $\mu$w, while 10 s for the excitation powers of 20 $\mu$w and 3 $\mu$w. Inset: QD level structure under a magnetic field with Faraday configuration, with circularly polarized exciton transitions $\sigma^+$ and $\sigma^-$.}
\end{figure*}
To investigate the coupling between a waveguide mode and a single QD in the slow light regime, we select the waveguide mode C close to the band edge from PCW4 with single QDs nearby, as shown in Fig. 3(a). Figure 3(a) presents the power-dependent PL spectra of the waveguide mode C and QDs at 5 K without normalization. It is noted that more waveguide modes exist within the wavelength range of Fig. 3(a), as the spectra obtained at higher excitation power in Fig. 1(e) indicate. Those waveguide modes are not visible due to the low excitation power used for the experiment in Fig. 3. When the excitation power increases, the waveguide mode C hardly shifts due to the minor change in refractive index caused by the thermo-optic effect or the carrier concentration increase. However, the emission energies of single QDs are more sensitive to the variation of the excitation power as the temperature increases simultaneously, leading to the bandgap shrinkage of QDs. Therefore, the waveguide mode C and a single quantum dot (QD1) can be clearly identified by the power-dependent PL spectra in Fig. 3(a). The wavelength variation of QD1 induced directly by the excitation power change of 0.2 $\mu$w to 100 $\mu$w is about 1.3 nm. The power control offers a simple and effective approach to tuning the PL energy of the target single quantum dot. The 1.3 nm shift of QD1 emission wavelength tuned by the excitation power is approximately equivalent to the wavelength shift by heating the whole sample from 4.2 K to about 52 K. The waveguide mode C is fitted with a Q factor of about 928, as shown by the red line in Fig. 3(b), and a group index of about 19 extracted from Fig. 1(e), which is 2.7 times of the group index in the linear dispersion region. Figure 3(b) shows that several PL peaks from QDs are close to the waveguide mode C. The wavelength of QDs can be easily tuned by changing the optical excitation power and the magnetic field, enabling the investigation of the interaction between a waveguide mode and single QDs.
%Due to thermo-optic effect, the refractive index of the device changes with varying temperature.

\begin{figure}[htp]
	\includegraphics[width=8.6 cm]{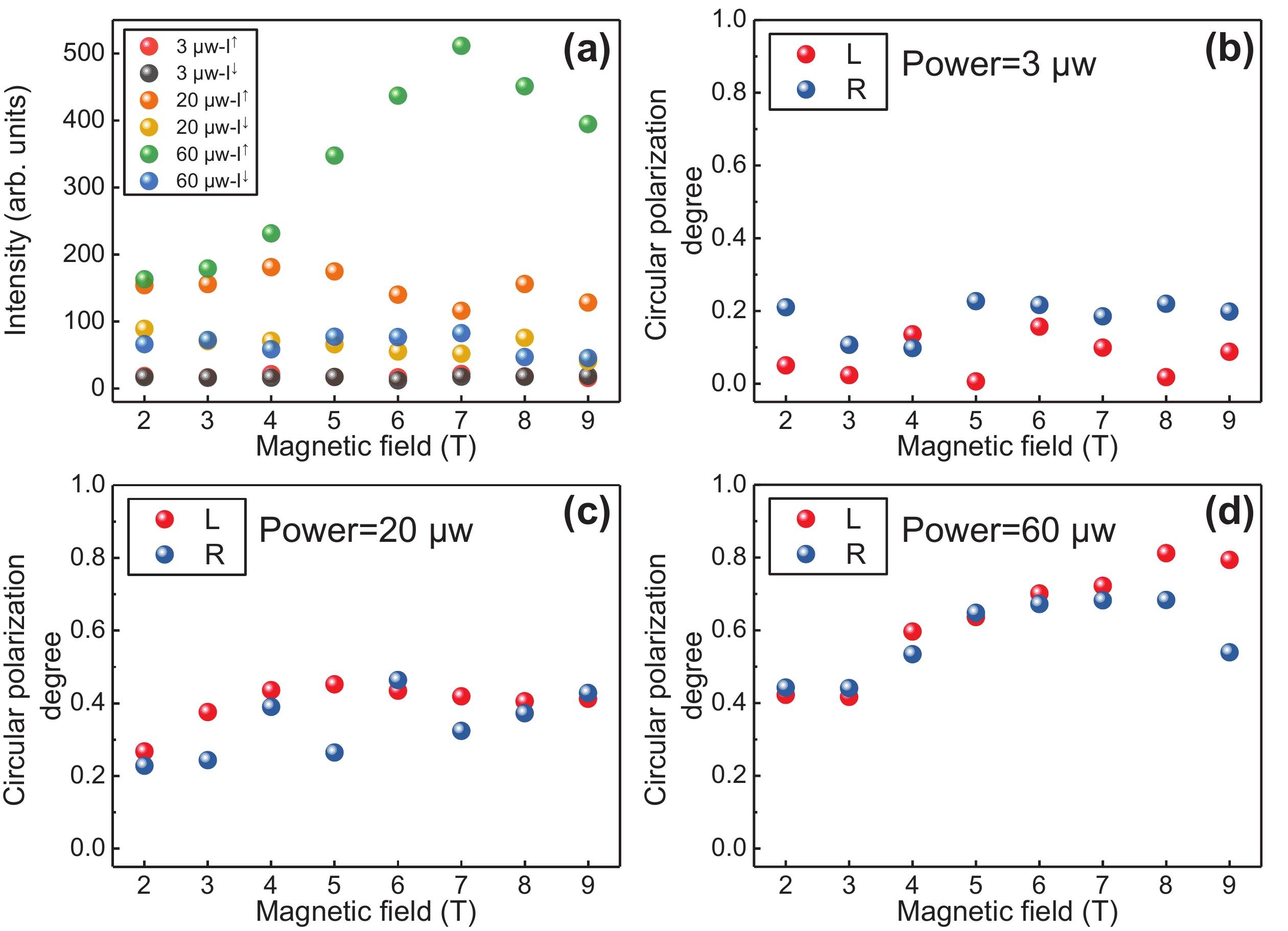}
	\caption{(a) The intensity variation of X$^\uparrow$ and X$^\downarrow$ emissions from QD1 tuned by the magnetic field under different excitation powers. The intensities are obtained by Lorentzian fitting of the PL spectra collected from the left grating coupler in Fig. 4. Magnetic field dependent circular polarization degrees of QD1 obtained by Lorentzian fitting of the PL spectra collected from the left (red) and right (blue) grating coupler with different excitation powers of (b) 3  $\mu$w, (c) 20 $\mu$w and (d) 60 $\mu$w in Fig. 4. }
\end{figure}
The exciton transition of a single QD is presented in the inset of Fig. 4(d) with an applied Faraday configuration magnetic field in parallel with the QDs’ growth direction. It is noted that the dark states resulting from mixing bright and dark-exciton states of QDs under the magnetic field in Voigt configuration are absent since the magnetic field is in Faraday configuration in this work \cite{peng2017probing}. The exciton is constructed by an electron with spin $S_e=\pm1/2$ and a hole with spin $J_h= \pm3/2$  \cite{kim2011magnetic}. States with angular momentum projections $M=\pm1$ are optical active under the Faraday configuration magnetic field, where ${M=S_e+J_h}$ \cite{bayer2002fine}. The magnetic field applied on a single QD, resulting in two spin states with opposite circular polarizations through the Zeeman effect \cite{kuther1998zeeman}, as shown in the inset in Fig. 4(d). The two peaks are labeled with X$^\uparrow$ and X$^\downarrow$. The Zeeman splitting between X$^\uparrow$ and X$^\downarrow$ is described with the equation ${\Delta}E{_{Zeeman}}=g{\mu_B}B$, where $\mu_B$ is the Bohr magneton and $g$ is the $g$ factor of the exciton. In addition, the two spin states also have an energy shift due to the diamagnetic effect \cite{cao2016observation,peng2017probing}, which is ${\Delta}E{_{dia}}={\gamma}B^2$, where $\gamma$ denotes the diamagnetic coefficient. The $g$ factor and the diamagnetic coefficient extracted from Fig. 4(a) are 3.9 and 5.7 $\mu$eV/T$^2$, respectively. By increasing the magnetic field from 0 T to 9 T, as shown in Fig. 4(a), the emission wavelength of X$^\uparrow$ is redshifted about 1 nm, while that of X$^\downarrow$ is blueshifted about 0.4 nm. The opposite energy shifts of X$^\uparrow$ and X$^\downarrow$ under the magnetic field can be used to investigate the two exciton states seperately coupling with a waveguide mode.

The magneto-optical investigation on the resonance between the single quantum dot QD1 and the waveguide mode C in PCW4 is performed to achieve a tunable spin-resolved enhancement transmission. Figures 4(a)-4(c) show the PL transmission spectra of QD1 coupling with the waveguide mode C collected through the left (red lines) and right (black lines) grating couplers of PCW4 under different excitation powers of 3 $\mu$w, 20 $\mu$w and 60 $\mu$w from 0 T to 9 T, respectively. The linewidths of the observed peaks at 0 T in Fig. 4(a)-4(c) are near 0.074 nm without obvious broadening. The waveguide mode C is marked by the orange area in Fig. 4(a)-4(c). Apparently, PL peaks with different spin states are enhanced when overlapping with the waveguide mode. With each excitation power, the overlapping degree between the polarized emissions and the waveguide mode C is also tuned by a magnetic field. Figure 4(d) shows the enhanced X$^\uparrow$ and X$^\downarrow$ peaks of QD1 collected from the left grating coupler under each excitation power with a magnetic field of 7 T. The selective enhancement becomes more pronounced when only one branch of the emissions of X$^\uparrow$ and X$^\downarrow$ is coupled with the waveguide mode. For instance, 60 $\mu$w will be ideal for the coupled system of QD1 and PCW4 to obtain a high circularly polarized degree, as shown in Fig. 4(d).
% And the intensity ratio of the emissions from X$^\uparrow$ and X$^\downarrow$ is demonstrated to be optically controllable without additional complex adjustment.

To quantify the intensity ratio of the spin-resolved emissions with opposite circular polarizations from X$^\uparrow$ and X$^\downarrow$ of QD1 coupled with the waveguide mode C, we define the circular polarization degree as $P_c$=($I^\uparrow$-$I^\downarrow$)/($I^\uparrow$+$I^\downarrow$), where I$^\uparrow$ (I$^\downarrow$) denotes the PL intensity of X$^\uparrow$ (X$^\downarrow$) peak. Different from the chiral contrast used in references \cite{sollner2015deterministic,coles2016chirality,xiao2021position,xiao2021chiral} with similar equation form, which describes the chiral coupling, $P_c$ represents the intensity contrast of spin state emissions with corresponding circular polarizations \cite{bayer2002fine}. Figure 5(a) shows the intensity variation of X$^\uparrow$ and X$^\downarrow$ emissions tuned by the magnetic field, corresponding to I$^\uparrow$ and I$^\downarrow$. I$^\uparrow$ and I$^\downarrow$ are extracted from spectra collected from the left grating coupler in Fig. 4(a)-4(c) under different excitation powers of 3 $\mu$w, 20 $\mu$w and 60 $\mu$w. I$^\uparrow$ and I$^\downarrow$ are almost equal with an excitation power of 3 $\mu$w. However, I$^\uparrow$ is enhanced by approximately 9 times compared with I$^\downarrow$ under B = 8 T with an excitation power of 60 $\mu$w.

The magnetic field dependent circular polarization degrees with 3 $\mu$w, 20 $\mu$w, and 60 $\mu$w are shown in Fig. 5(b)-5(c), which are extracted by Lorentzian fitting of the intensity of each peak from PL spectra in Fig. 4(a)-4(c). The circular polarization degree is generally low under an excitation power of 3 $\mu$w. The random distribution from 0 to 0.23 of circular polarization degrees indicates that the emission peaks of X$^\uparrow$ and X$^\downarrow$ are both coupled with the waveguide mode C to the same extent, as shown in Fig. 4(a). When the excitation power reaches 20 $\mu$w, the circular polarization degree gradually increases from 0.17 to 0.43. The different degree of enhancement for two peaks of X$^\uparrow$ and X$^\downarrow$ results from the unequal overlapping between two peaks and the waveguide mode C (see Fig. 4(b)). At the power of 60 $\mu$w, the emission wavelength of QD1 at 0 T locates at the edge of the waveguide mode C, as shown in Fig. 4(c). The spin state X$^\downarrow$ is not coupled with waveguide mode as it shifts away from the waveguide mode C with the increasing magnetic field, while X$^\uparrow$ moves towards the waveguide mode C with a better spectral overlapping. A maximum circular polarization degree of 0.81 is obtained at 8 T, indicating the spin state X$^\uparrow$ can be selectively coupled with the waveguide mode C. The polarization degree decreases at 9 T because the spin state X$^\uparrow$ is away from the center wavelength of the waveguide mode. In principle, the circular polarization degree can be further increased by optimizing the design and fabrication of PCWs. For instance, the site-control of QDs could also be applied in PCWs \cite{jarlov2015exciton,rigal2018single}, which enables better spatial overlap between QDs and field antinodes of waveguides for stronger coupling \cite{rao2007single}.

\section{\label{sec:level4}Conclusion}
We demonstrated a controllable enhancement of single QD exciton spin states coupling to a slow-light mode in a PCW under a magnetic field. PCWs with slow light modes have been simulated and experimentally demonstrated to match the luminescence of QDs. Two spin states are separately coupled to the waveguide mode by tuning the magnetic field and optical excitation pumping power. The intensity contrast of spin state emissions up to 0.81 is achieved when one of the spin states couples with the slow-light mode effectively while the other is tuned away from the waveguide mode. The deterministic high circular polarization degree benefits the realization of circularly polarized photon emissions. Such a tunable in-plane spin-polarized photon source will be helpful for on-chip scalable quantum optical circuits \cite{chen2021quantum} for future information processing.

\begin{acknowledgments}
This work was supported by the National Key Research and Development Program of China (Grant No. 2021YFA1400700), the National Natural Science Foundation of China (Grants Nos. 62025507, 11934019, 92250301, 11721404, 62175254 and 12204020), the Strategic Priority Research Program (Grant No. XDB28000000) of the Chinese Academy of Sciences.
\end{acknowledgments}
%apsrev4-2.bst 2019-01-14 (MD) hand-edited version of apsrev4-1.bst
%Control: key (0)
%Control: author (8) initials jnrlst
%Control: editor formatted (1) identically to author
%Control: production of article title (0) allowed
%Control: page (0) single
%Control: year (1) truncated
%Control: production of eprint (0) enabled
%
% Produces the bibliography via BibTeX.

\begin{thebibliography}{51}%
\makeatletter
\providecommand \@ifxundefined [1]{%
 \@ifx{#1\undefined}
}%
\providecommand \@ifnum [1]{%
 \ifnum #1\expandafter \@firstoftwo
 \else \expandafter \@secondoftwo
 \fi
}%
\providecommand \@ifx [1]{%
 \ifx #1\expandafter \@firstoftwo
 \else \expandafter \@secondoftwo
 \fi
}%
\providecommand \natexlab [1]{#1}%
\providecommand \enquote  [1]{``#1''}%
\providecommand \bibnamefont  [1]{#1}%
\providecommand \bibfnamefont [1]{#1}%
\providecommand \citenamefont [1]{#1}%
\providecommand \href@noop [0]{\@secondoftwo}%
\providecommand \href [0]{\begingroup \@sanitize@url \@href}%
\providecommand \@href[1]{\@@startlink{#1}\@@href}%
\providecommand \@@href[1]{\endgroup#1\@@endlink}%
\providecommand \@sanitize@url [0]{\catcode `\\12\catcode `\$12\catcode
  `\&12\catcode `\#12\catcode `\^12\catcode `\_12\catcode `\%12\relax}%
\providecommand \@@startlink[1]{}%
\providecommand \@@endlink[0]{}%
\providecommand \url  [0]{\begingroup\@sanitize@url \@url }%
\providecommand \@url [1]{\endgroup\@href {#1}{\urlprefix }}%
\providecommand \urlprefix  [0]{URL }%
\providecommand \Eprint [0]{\href }%
\providecommand \doibase [0]{https://doi.org/}%
\providecommand \selectlanguage [0]{\@gobble}%
\providecommand \bibinfo  [0]{\@secondoftwo}%
\providecommand \bibfield  [0]{\@secondoftwo}%
\providecommand \translation [1]{[#1]}%
\providecommand \BibitemOpen [0]{}%
\providecommand \bibitemStop [0]{}%
\providecommand \bibitemNoStop [0]{.\EOS\space}%
\providecommand \EOS [0]{\spacefactor3000\relax}%
\providecommand \BibitemShut  [1]{\csname bibitem#1\endcsname}%
\let\auto@bib@innerbib\@empty
%</preamble>
\bibitem [{\citenamefont {Lodahl}\ \emph {et~al.}(2015)\citenamefont {Lodahl},
  \citenamefont {Mahmoodian},\ and\ \citenamefont
  {Stobbe}}]{lodahl2015interfacing}%
  \BibitemOpen
  \bibfield  {author} {\bibinfo {author} {\bibfnamefont {P.}~\bibnamefont
  {Lodahl}}, \bibinfo {author} {\bibfnamefont {S.}~\bibnamefont {Mahmoodian}},\
  and\ \bibinfo {author} {\bibfnamefont {S.}~\bibnamefont {Stobbe}},\
  }\bibfield  {title} {\bibinfo {title} {Interfacing single photons and single
  quantum dots with photonic nanostructures},\ }\href@noop {} {\bibfield
  {journal} {\bibinfo  {journal} {Rev. Mod. Phys.}\ }\textbf {\bibinfo {volume}
  {87}},\ \bibinfo {pages} {347} (\bibinfo {year} {2015})}\BibitemShut
  {NoStop}%
\bibitem [{\citenamefont {Dietrich}\ \emph {et~al.}(2016)\citenamefont
  {Dietrich}, \citenamefont {Fiore}, \citenamefont {Thompson}, \citenamefont
  {Kamp},\ and\ \citenamefont {Hofling}}]{dietrich2016gaas}%
  \BibitemOpen
  \bibfield  {author} {\bibinfo {author} {\bibfnamefont {C.~P.}\ \bibnamefont
  {Dietrich}}, \bibinfo {author} {\bibfnamefont {A.}~\bibnamefont {Fiore}},
  \bibinfo {author} {\bibfnamefont {M.~G.}\ \bibnamefont {Thompson}}, \bibinfo
  {author} {\bibfnamefont {M.}~\bibnamefont {Kamp}},\ and\ \bibinfo {author}
  {\bibfnamefont {S.}~\bibnamefont {Hofling}},\ }\bibfield  {title} {\bibinfo
  {title} {Gaas integrated quantum photonics: Towards compact and
  multi-functional quantum photonic integrated circuits},\ }\href@noop {}
  {\bibfield  {journal} {\bibinfo  {journal} {Laser Photonics Rev.}\ }\textbf
  {\bibinfo {volume} {10}},\ \bibinfo {pages} {870} (\bibinfo {year}
  {2016})}\BibitemShut {NoStop}%
\bibitem [{\citenamefont {Senellart}\ \emph {et~al.}(2017)\citenamefont
  {Senellart}, \citenamefont {Solomon},\ and\ \citenamefont
  {White}}]{senellart2017high}%
  \BibitemOpen
  \bibfield  {author} {\bibinfo {author} {\bibfnamefont {P.}~\bibnamefont
  {Senellart}}, \bibinfo {author} {\bibfnamefont {G.}~\bibnamefont {Solomon}},\
  and\ \bibinfo {author} {\bibfnamefont {A.}~\bibnamefont {White}},\ }\bibfield
   {title} {\bibinfo {title} {High-performance semiconductor quantum-dot
  single-photon sources},\ }\href@noop {} {\bibfield  {journal} {\bibinfo
  {journal} {Nat. Nanotechnol.}\ }\textbf {\bibinfo {volume} {12}},\ \bibinfo
  {pages} {1026} (\bibinfo {year} {2017})}\BibitemShut {NoStop}%
\bibitem [{\citenamefont {Arakawa}\ and\ \citenamefont
  {Holmes}(2020)}]{arakawa2020progress}%
  \BibitemOpen
  \bibfield  {author} {\bibinfo {author} {\bibfnamefont {Y.}~\bibnamefont
  {Arakawa}}\ and\ \bibinfo {author} {\bibfnamefont {M.~J.}\ \bibnamefont
  {Holmes}},\ }\bibfield  {title} {\bibinfo {title} {Progress in quantum-dot
  single photon sources for quantum information technologies: A broad spectrum
  overview},\ }\href@noop {} {\bibfield  {journal} {\bibinfo  {journal} {Appl.
  Phys. Rev.}\ }\textbf {\bibinfo {volume} {7}},\ \bibinfo {pages} {021309}
  (\bibinfo {year} {2020})}\BibitemShut {NoStop}%
\bibitem [{\citenamefont {Santori}\ \emph {et~al.}(2002)\citenamefont
  {Santori}, \citenamefont {Fattal}, \citenamefont {Vu{\v{c}}kovi{\'c}},
  \citenamefont {Solomon},\ and\ \citenamefont
  {Yamamoto}}]{santori2002indistinguishable}%
  \BibitemOpen
  \bibfield  {author} {\bibinfo {author} {\bibfnamefont {C.}~\bibnamefont
  {Santori}}, \bibinfo {author} {\bibfnamefont {D.}~\bibnamefont {Fattal}},
  \bibinfo {author} {\bibfnamefont {J.}~\bibnamefont {Vu{\v{c}}kovi{\'c}}},
  \bibinfo {author} {\bibfnamefont {G.~S.}\ \bibnamefont {Solomon}},\ and\
  \bibinfo {author} {\bibfnamefont {Y.}~\bibnamefont {Yamamoto}},\ }\bibfield
  {title} {\bibinfo {title} {Indistinguishable photons from a single-photon
  device},\ }\href@noop {} {\bibfield  {journal} {\bibinfo  {journal} {Nature}\
  }\textbf {\bibinfo {volume} {419}},\ \bibinfo {pages} {594} (\bibinfo {year}
  {2002})}\BibitemShut {NoStop}%
\bibitem [{\citenamefont {Pelton}\ \emph {et~al.}(2002)\citenamefont {Pelton},
  \citenamefont {Santori}, \citenamefont {Vuckovi{\'c}}, \citenamefont {Zhang},
  \citenamefont {Solomon}, \citenamefont {Plant},\ and\ \citenamefont
  {Yamamoto}}]{pelton2002efficient}%
  \BibitemOpen
  \bibfield  {author} {\bibinfo {author} {\bibfnamefont {M.}~\bibnamefont
  {Pelton}}, \bibinfo {author} {\bibfnamefont {C.}~\bibnamefont {Santori}},
  \bibinfo {author} {\bibfnamefont {J.}~\bibnamefont {Vuckovi{\'c}}}, \bibinfo
  {author} {\bibfnamefont {B.}~\bibnamefont {Zhang}}, \bibinfo {author}
  {\bibfnamefont {G.~S.}\ \bibnamefont {Solomon}}, \bibinfo {author}
  {\bibfnamefont {J.}~\bibnamefont {Plant}},\ and\ \bibinfo {author}
  {\bibfnamefont {Y.}~\bibnamefont {Yamamoto}},\ }\bibfield  {title} {\bibinfo
  {title} {Efficient source of single photons: a single quantum dot in a
  micropost microcavity},\ }\href@noop {} {\bibfield  {journal} {\bibinfo
  {journal} {Phys. Rev. Lett.}\ }\textbf {\bibinfo {volume} {89}},\ \bibinfo
  {pages} {233602} (\bibinfo {year} {2002})}\BibitemShut {NoStop}%
\bibitem [{\citenamefont {Gazzano}\ \emph {et~al.}(2013)\citenamefont
  {Gazzano}, \citenamefont {Michaelis~de Vasconcellos}, \citenamefont {Arnold},
  \citenamefont {Nowak}, \citenamefont {Galopin}, \citenamefont {Sagnes},
  \citenamefont {Lanco}, \citenamefont {Lema{\^\i}tre},\ and\ \citenamefont
  {Senellart}}]{gazzano2013bright}%
  \BibitemOpen
  \bibfield  {author} {\bibinfo {author} {\bibfnamefont {O.}~\bibnamefont
  {Gazzano}}, \bibinfo {author} {\bibfnamefont {S.}~\bibnamefont {Michaelis~de
  Vasconcellos}}, \bibinfo {author} {\bibfnamefont {C.}~\bibnamefont {Arnold}},
  \bibinfo {author} {\bibfnamefont {A.}~\bibnamefont {Nowak}}, \bibinfo
  {author} {\bibfnamefont {E.}~\bibnamefont {Galopin}}, \bibinfo {author}
  {\bibfnamefont {I.}~\bibnamefont {Sagnes}}, \bibinfo {author} {\bibfnamefont
  {L.}~\bibnamefont {Lanco}}, \bibinfo {author} {\bibfnamefont
  {A.}~\bibnamefont {Lema{\^\i}tre}},\ and\ \bibinfo {author} {\bibfnamefont
  {P.}~\bibnamefont {Senellart}},\ }\bibfield  {title} {\bibinfo {title}
  {Bright solid-state sources of indistinguishable single photons},\
  }\href@noop {} {\bibfield  {journal} {\bibinfo  {journal} {Nat. Commun.}\
  }\textbf {\bibinfo {volume} {4}},\ \bibinfo {pages} {1425} (\bibinfo {year}
  {2013})}\BibitemShut {NoStop}%
\bibitem [{\citenamefont {Schlehahn}\ \emph {et~al.}(2016)\citenamefont
  {Schlehahn}, \citenamefont {Thoma}, \citenamefont {Munnelly}, \citenamefont
  {Kamp}, \citenamefont {H{\"o}fling}, \citenamefont {Heindel}, \citenamefont
  {Schneider},\ and\ \citenamefont {Reitzenstein}}]{schlehahn2016electrically}%
  \BibitemOpen
  \bibfield  {author} {\bibinfo {author} {\bibfnamefont {A.}~\bibnamefont
  {Schlehahn}}, \bibinfo {author} {\bibfnamefont {A.}~\bibnamefont {Thoma}},
  \bibinfo {author} {\bibfnamefont {P.}~\bibnamefont {Munnelly}}, \bibinfo
  {author} {\bibfnamefont {M.}~\bibnamefont {Kamp}}, \bibinfo {author}
  {\bibfnamefont {S.}~\bibnamefont {H{\"o}fling}}, \bibinfo {author}
  {\bibfnamefont {T.}~\bibnamefont {Heindel}}, \bibinfo {author} {\bibfnamefont
  {C.}~\bibnamefont {Schneider}},\ and\ \bibinfo {author} {\bibfnamefont
  {S.}~\bibnamefont {Reitzenstein}},\ }\bibfield  {title} {\bibinfo {title} {An
  electrically driven cavity-enhanced source of indistinguishable photons with
  61\% overall efficiency},\ }\href@noop {} {\bibfield  {journal} {\bibinfo
  {journal} {APL Photonics}\ }\textbf {\bibinfo {volume} {1}},\ \bibinfo
  {pages} {011301} (\bibinfo {year} {2016})}\BibitemShut {NoStop}%
\bibitem [{\citenamefont {Kiraz}\ \emph {et~al.}(2001)\citenamefont {Kiraz},
  \citenamefont {Michler}, \citenamefont {Becher}, \citenamefont {Gayral},
  \citenamefont {Imamo{\u{g}}lu}, \citenamefont {Zhang}, \citenamefont {Hu},
  \citenamefont {Schoenfeld},\ and\ \citenamefont {Petroff}}]{kiraz2001cavity}%
  \BibitemOpen
  \bibfield  {author} {\bibinfo {author} {\bibfnamefont {A.}~\bibnamefont
  {Kiraz}}, \bibinfo {author} {\bibfnamefont {P.}~\bibnamefont {Michler}},
  \bibinfo {author} {\bibfnamefont {C.}~\bibnamefont {Becher}}, \bibinfo
  {author} {\bibfnamefont {B.}~\bibnamefont {Gayral}}, \bibinfo {author}
  {\bibfnamefont {A.}~\bibnamefont {Imamo{\u{g}}lu}}, \bibinfo {author}
  {\bibfnamefont {L.}~\bibnamefont {Zhang}}, \bibinfo {author} {\bibfnamefont
  {E.}~\bibnamefont {Hu}}, \bibinfo {author} {\bibfnamefont {W.}~\bibnamefont
  {Schoenfeld}},\ and\ \bibinfo {author} {\bibfnamefont {P.}~\bibnamefont
  {Petroff}},\ }\bibfield  {title} {\bibinfo {title} {Cavity-quantum
  electrodynamics using a single inas quantum dot in a microdisk structure},\
  }\href@noop {} {\bibfield  {journal} {\bibinfo  {journal} {Appl. Phys.
  Lett.}\ }\textbf {\bibinfo {volume} {78}},\ \bibinfo {pages} {3932} (\bibinfo
  {year} {2001})}\BibitemShut {NoStop}%
\bibitem [{\citenamefont {Yang}\ \emph {et~al.}(2021)\citenamefont {Yang},
  \citenamefont {Shi}, \citenamefont {Xie}, \citenamefont {Wu}, \citenamefont
  {Xiao}, \citenamefont {Song}, \citenamefont {Dang}, \citenamefont {Sun},
  \citenamefont {Yang}, \citenamefont {Ge}, \citenamefont {Li}, \citenamefont
  {Zuo}, \citenamefont {Jin},\ and\ \citenamefont {Xu}}]{yang2021enhanced}%
  \BibitemOpen
  \bibfield  {author} {\bibinfo {author} {\bibfnamefont {J.}~\bibnamefont
  {Yang}}, \bibinfo {author} {\bibfnamefont {S.}~\bibnamefont {Shi}}, \bibinfo
  {author} {\bibfnamefont {X.}~\bibnamefont {Xie}}, \bibinfo {author}
  {\bibfnamefont {S.}~\bibnamefont {Wu}}, \bibinfo {author} {\bibfnamefont
  {S.}~\bibnamefont {Xiao}}, \bibinfo {author} {\bibfnamefont {F.}~\bibnamefont
  {Song}}, \bibinfo {author} {\bibfnamefont {J.}~\bibnamefont {Dang}}, \bibinfo
  {author} {\bibfnamefont {S.}~\bibnamefont {Sun}}, \bibinfo {author}
  {\bibfnamefont {L.}~\bibnamefont {Yang}}, \bibinfo {author} {\bibfnamefont
  {Z.-Y.}\ \bibnamefont {Ge}}, \bibinfo {author} {\bibfnamefont {B.-B.}\
  \bibnamefont {Li}}, \bibinfo {author} {\bibfnamefont {Z.}~\bibnamefont
  {Zuo}}, \bibinfo {author} {\bibfnamefont {K.}~\bibnamefont {Jin}},\ and\
  \bibinfo {author} {\bibfnamefont {X.}~\bibnamefont {Xu}},\ }\bibfield
  {title} {\bibinfo {title} {Enhanced emission from a single quantum dot in a
  microdisk at a deterministic diabolical point},\ }\href@noop {} {\bibfield
  {journal} {\bibinfo  {journal} {Opt. Express}\ }\textbf {\bibinfo {volume}
  {29}},\ \bibinfo {pages} {14231} (\bibinfo {year} {2021})}\BibitemShut
  {NoStop}%
\bibitem [{\citenamefont {Dusanowski}\ \emph {et~al.}(2020)\citenamefont
  {Dusanowski}, \citenamefont {Kock}, \citenamefont {Shin}, \citenamefont
  {Kwon}, \citenamefont {Schneider},\ and\ \citenamefont
  {Hofling}}]{dusanowski2020purcell}%
  \BibitemOpen
  \bibfield  {author} {\bibinfo {author} {\bibfnamefont {{\L}.}~\bibnamefont
  {Dusanowski}}, \bibinfo {author} {\bibfnamefont {D.}~\bibnamefont {Kock}},
  \bibinfo {author} {\bibfnamefont {E.}~\bibnamefont {Shin}}, \bibinfo {author}
  {\bibfnamefont {S.-H.}\ \bibnamefont {Kwon}}, \bibinfo {author}
  {\bibfnamefont {C.}~\bibnamefont {Schneider}},\ and\ \bibinfo {author}
  {\bibfnamefont {S.}~\bibnamefont {Hofling}},\ }\bibfield  {title} {\bibinfo
  {title} {Purcell-enhanced and indistinguishable single-photon generation from
  quantum dots coupled to on-chip integrated ring resonators},\ }\href@noop {}
  {\bibfield  {journal} {\bibinfo  {journal} {Nano Lett.}\ }\textbf {\bibinfo
  {volume} {20}},\ \bibinfo {pages} {6357} (\bibinfo {year}
  {2020})}\BibitemShut {NoStop}%
\bibitem [{\citenamefont {Kir{\v{s}}ansk{\.e}}\ \emph
  {et~al.}(2017)\citenamefont {Kir{\v{s}}ansk{\.e}}, \citenamefont
  {Thyrrestrup}, \citenamefont {Daveau}, \citenamefont {Dree{\ss}en},
  \citenamefont {Pregnolato}, \citenamefont {Midolo}, \citenamefont
  {Tighineanu}, \citenamefont {Javadi}, \citenamefont {Stobbe}, \citenamefont
  {Schott}, \citenamefont {Ludwig}, \citenamefont {Wieck}, \citenamefont
  {Park}, \citenamefont {Song}, \citenamefont {Kuhlmann}, \citenamefont
  {S\"ollner}, \citenamefont {L\"obl}, \citenamefont {Warburton},\ and\
  \citenamefont {Lodahl}}]{kirvsanske2017indistinguishable}%
  \BibitemOpen
  \bibfield  {author} {\bibinfo {author} {\bibfnamefont {G.}~\bibnamefont
  {Kir{\v{s}}ansk{\.e}}}, \bibinfo {author} {\bibfnamefont {H.}~\bibnamefont
  {Thyrrestrup}}, \bibinfo {author} {\bibfnamefont {R.~S.}\ \bibnamefont
  {Daveau}}, \bibinfo {author} {\bibfnamefont {C.~L.}\ \bibnamefont
  {Dree{\ss}en}}, \bibinfo {author} {\bibfnamefont {T.}~\bibnamefont
  {Pregnolato}}, \bibinfo {author} {\bibfnamefont {L.}~\bibnamefont {Midolo}},
  \bibinfo {author} {\bibfnamefont {P.}~\bibnamefont {Tighineanu}}, \bibinfo
  {author} {\bibfnamefont {A.}~\bibnamefont {Javadi}}, \bibinfo {author}
  {\bibfnamefont {S.}~\bibnamefont {Stobbe}}, \bibinfo {author} {\bibfnamefont
  {R.}~\bibnamefont {Schott}}, \bibinfo {author} {\bibfnamefont
  {A.}~\bibnamefont {Ludwig}}, \bibinfo {author} {\bibfnamefont {A.~D.}\
  \bibnamefont {Wieck}}, \bibinfo {author} {\bibfnamefont {S.~I.}\ \bibnamefont
  {Park}}, \bibinfo {author} {\bibfnamefont {J.~D.}\ \bibnamefont {Song}},
  \bibinfo {author} {\bibfnamefont {A.~V.}\ \bibnamefont {Kuhlmann}}, \bibinfo
  {author} {\bibfnamefont {I.}~\bibnamefont {S\"ollner}}, \bibinfo {author}
  {\bibfnamefont {M.~C.}\ \bibnamefont {L\"obl}}, \bibinfo {author}
  {\bibfnamefont {R.~J.}\ \bibnamefont {Warburton}},\ and\ \bibinfo {author}
  {\bibfnamefont {P.}~\bibnamefont {Lodahl}},\ }\bibfield  {title} {\bibinfo
  {title} {Indistinguishable and efficient single photons from a quantum dot in
  a planar nanobeam waveguide},\ }\href@noop {} {\bibfield  {journal} {\bibinfo
   {journal} {Phys. Rev. B}\ }\textbf {\bibinfo {volume} {96}},\ \bibinfo
  {pages} {165306} (\bibinfo {year} {2017})}\BibitemShut {NoStop}%
\bibitem [{\citenamefont {Hepp}\ \emph {et~al.}(2020)\citenamefont {Hepp},
  \citenamefont {Hornung}, \citenamefont {Bauer}, \citenamefont {Hesselmeier},
  \citenamefont {Yuan}, \citenamefont {Jetter}, \citenamefont {Portalupi},
  \citenamefont {Rastelli},\ and\ \citenamefont {Michler}}]{hepp2020purcell}%
  \BibitemOpen
  \bibfield  {author} {\bibinfo {author} {\bibfnamefont {S.}~\bibnamefont
  {Hepp}}, \bibinfo {author} {\bibfnamefont {F.}~\bibnamefont {Hornung}},
  \bibinfo {author} {\bibfnamefont {S.}~\bibnamefont {Bauer}}, \bibinfo
  {author} {\bibfnamefont {E.}~\bibnamefont {Hesselmeier}}, \bibinfo {author}
  {\bibfnamefont {X.}~\bibnamefont {Yuan}}, \bibinfo {author} {\bibfnamefont
  {M.}~\bibnamefont {Jetter}}, \bibinfo {author} {\bibfnamefont {S.~L.}\
  \bibnamefont {Portalupi}}, \bibinfo {author} {\bibfnamefont {A.}~\bibnamefont
  {Rastelli}},\ and\ \bibinfo {author} {\bibfnamefont {P.}~\bibnamefont
  {Michler}},\ }\bibfield  {title} {\bibinfo {title} {Purcell-enhanced
  single-photon emission from a strain-tunable quantum dot in a
  cavity-waveguide device},\ }\href@noop {} {\bibfield  {journal} {\bibinfo
  {journal} {Appl. Phys. Lett.}\ }\textbf {\bibinfo {volume} {117}},\ \bibinfo
  {pages} {254002} (\bibinfo {year} {2020})}\BibitemShut {NoStop}%
\bibitem [{\citenamefont {Gong}\ \emph {et~al.}(2010)\citenamefont {Gong},
  \citenamefont {Ellis}, \citenamefont {Shambat}, \citenamefont {Sarmiento},
  \citenamefont {Harris},\ and\ \citenamefont
  {Vu{\v{c}}kovi{\'c}}}]{gong2010nanobeam}%
  \BibitemOpen
  \bibfield  {author} {\bibinfo {author} {\bibfnamefont {Y.}~\bibnamefont
  {Gong}}, \bibinfo {author} {\bibfnamefont {B.}~\bibnamefont {Ellis}},
  \bibinfo {author} {\bibfnamefont {G.}~\bibnamefont {Shambat}}, \bibinfo
  {author} {\bibfnamefont {T.}~\bibnamefont {Sarmiento}}, \bibinfo {author}
  {\bibfnamefont {J.~S.}\ \bibnamefont {Harris}},\ and\ \bibinfo {author}
  {\bibfnamefont {J.}~\bibnamefont {Vu{\v{c}}kovi{\'c}}},\ }\bibfield  {title}
  {\bibinfo {title} {Nanobeam photonic crystal cavity quantum dot laser},\
  }\href@noop {} {\bibfield  {journal} {\bibinfo  {journal} {Opt. Express}\
  }\textbf {\bibinfo {volume} {18}},\ \bibinfo {pages} {8781} (\bibinfo {year}
  {2010})}\BibitemShut {NoStop}%
\bibitem [{\citenamefont {Englund}\ \emph {et~al.}(2005)\citenamefont
  {Englund}, \citenamefont {Fattal}, \citenamefont {Waks}, \citenamefont
  {Solomon}, \citenamefont {Zhang}, \citenamefont {Nakaoka}, \citenamefont
  {Arakawa}, \citenamefont {Yamamoto},\ and\ \citenamefont
  {Vu{\v{c}}kovi{\'c}}}]{englund2005controlling}%
  \BibitemOpen
  \bibfield  {author} {\bibinfo {author} {\bibfnamefont {D.}~\bibnamefont
  {Englund}}, \bibinfo {author} {\bibfnamefont {D.}~\bibnamefont {Fattal}},
  \bibinfo {author} {\bibfnamefont {E.}~\bibnamefont {Waks}}, \bibinfo {author}
  {\bibfnamefont {G.}~\bibnamefont {Solomon}}, \bibinfo {author} {\bibfnamefont
  {B.}~\bibnamefont {Zhang}}, \bibinfo {author} {\bibfnamefont
  {T.}~\bibnamefont {Nakaoka}}, \bibinfo {author} {\bibfnamefont
  {Y.}~\bibnamefont {Arakawa}}, \bibinfo {author} {\bibfnamefont
  {Y.}~\bibnamefont {Yamamoto}},\ and\ \bibinfo {author} {\bibfnamefont
  {J.}~\bibnamefont {Vu{\v{c}}kovi{\'c}}},\ }\bibfield  {title} {\bibinfo
  {title} {Controlling the spontaneous emission rate of single quantum dots in
  a two-dimensional photonic crystal},\ }\href@noop {} {\bibfield  {journal}
  {\bibinfo  {journal} {Phys. Rev. Lett.}\ }\textbf {\bibinfo {volume} {95}},\
  \bibinfo {pages} {013904} (\bibinfo {year} {2005})}\BibitemShut {NoStop}%
\bibitem [{\citenamefont {Chang}\ \emph {et~al.}(2006)\citenamefont {Chang},
  \citenamefont {Chen}, \citenamefont {Chang}, \citenamefont {Hsieh},
  \citenamefont {Chyi},\ and\ \citenamefont {Hsu}}]{chang2006efficient}%
  \BibitemOpen
  \bibfield  {author} {\bibinfo {author} {\bibfnamefont {W.}~\bibnamefont
  {Chang}}, \bibinfo {author} {\bibfnamefont {W.}~\bibnamefont {Chen}},
  \bibinfo {author} {\bibfnamefont {H.}~\bibnamefont {Chang}}, \bibinfo
  {author} {\bibfnamefont {T.}~\bibnamefont {Hsieh}}, \bibinfo {author}
  {\bibfnamefont {J.}~\bibnamefont {Chyi}},\ and\ \bibinfo {author}
  {\bibfnamefont {T.}~\bibnamefont {Hsu}},\ }\bibfield  {title} {\bibinfo
  {title} {Efficient single-photon sources based on low-density quantum dots in
  photonic-crystal nanocavities},\ }\href@noop {} {\bibfield  {journal}
  {\bibinfo  {journal} {Phys. Rev. Lett.}\ }\textbf {\bibinfo {volume} {96}},\
  \bibinfo {pages} {117401} (\bibinfo {year} {2006})}\BibitemShut {NoStop}%
\bibitem [{\citenamefont {Balet}\ \emph {et~al.}(2007)\citenamefont {Balet},
  \citenamefont {Francardi}, \citenamefont {Gerardino}, \citenamefont
  {Chauvin}, \citenamefont {Alloing}, \citenamefont {Zinoni}, \citenamefont
  {Monat}, \citenamefont {Li}, \citenamefont {Le~Thomas}, \citenamefont
  {Houdré},\ and\ \citenamefont {Fiore}}]{balet2007enhanced}%
  \BibitemOpen
  \bibfield  {author} {\bibinfo {author} {\bibfnamefont {L.}~\bibnamefont
  {Balet}}, \bibinfo {author} {\bibfnamefont {M.}~\bibnamefont {Francardi}},
  \bibinfo {author} {\bibfnamefont {A.}~\bibnamefont {Gerardino}}, \bibinfo
  {author} {\bibfnamefont {N.}~\bibnamefont {Chauvin}}, \bibinfo {author}
  {\bibfnamefont {B.}~\bibnamefont {Alloing}}, \bibinfo {author} {\bibfnamefont
  {C.}~\bibnamefont {Zinoni}}, \bibinfo {author} {\bibfnamefont
  {C.}~\bibnamefont {Monat}}, \bibinfo {author} {\bibfnamefont {L.~H.}\
  \bibnamefont {Li}}, \bibinfo {author} {\bibfnamefont {N.}~\bibnamefont
  {Le~Thomas}}, \bibinfo {author} {\bibfnamefont {R.}~\bibnamefont {Houdré}},\
  and\ \bibinfo {author} {\bibfnamefont {A.}~\bibnamefont {Fiore}},\ }\bibfield
   {title} {\bibinfo {title} {Enhanced spontaneous emission rate from single
  inas quantum dots in a photonic crystal nanocavity at telecom wavelengths},\
  }\href@noop {} {\bibfield  {journal} {\bibinfo  {journal} {Appl. Phys.
  Lett.}\ }\textbf {\bibinfo {volume} {91}},\ \bibinfo {pages} {123115}
  (\bibinfo {year} {2007})}\BibitemShut {NoStop}%
\bibitem [{\citenamefont {Xie}\ \emph {et~al.}(2020)\citenamefont {Xie},
  \citenamefont {Zhang}, \citenamefont {He}, \citenamefont {Wu}, \citenamefont
  {Dang}, \citenamefont {Peng}, \citenamefont {Song}, \citenamefont {Yang},
  \citenamefont {Ni}, \citenamefont {Niu}, \citenamefont {Wang}, \citenamefont
  {Jin}, \citenamefont {Zhang},\ and\ \citenamefont {Xu}}]{xie2020cavity}%
  \BibitemOpen
  \bibfield  {author} {\bibinfo {author} {\bibfnamefont {X.}~\bibnamefont
  {Xie}}, \bibinfo {author} {\bibfnamefont {W.}~\bibnamefont {Zhang}}, \bibinfo
  {author} {\bibfnamefont {X.}~\bibnamefont {He}}, \bibinfo {author}
  {\bibfnamefont {S.}~\bibnamefont {Wu}}, \bibinfo {author} {\bibfnamefont
  {J.}~\bibnamefont {Dang}}, \bibinfo {author} {\bibfnamefont {K.}~\bibnamefont
  {Peng}}, \bibinfo {author} {\bibfnamefont {F.}~\bibnamefont {Song}}, \bibinfo
  {author} {\bibfnamefont {L.}~\bibnamefont {Yang}}, \bibinfo {author}
  {\bibfnamefont {H.}~\bibnamefont {Ni}}, \bibinfo {author} {\bibfnamefont
  {Z.}~\bibnamefont {Niu}}, \bibinfo {author} {\bibfnamefont {C.}~\bibnamefont
  {Wang}}, \bibinfo {author} {\bibfnamefont {K.}~\bibnamefont {Jin}}, \bibinfo
  {author} {\bibfnamefont {X.}~\bibnamefont {Zhang}},\ and\ \bibinfo {author}
  {\bibfnamefont {X.}~\bibnamefont {Xu}},\ }\bibfield  {title} {\bibinfo
  {title} {Cavity quantum electrodynamics with second-order topological corner
  state},\ }\href@noop {} {\bibfield  {journal} {\bibinfo  {journal} {Laser
  Photonics Rev.}\ }\textbf {\bibinfo {volume} {14}},\ \bibinfo {pages}
  {1900425} (\bibinfo {year} {2020})}\BibitemShut {NoStop}%
\bibitem [{\citenamefont {Yao}\ \emph {et~al.}(2010)\citenamefont {Yao},
  \citenamefont {Manga~Rao},\ and\ \citenamefont {Hughes}}]{yao2010chip}%
  \BibitemOpen
  \bibfield  {author} {\bibinfo {author} {\bibfnamefont {P.}~\bibnamefont
  {Yao}}, \bibinfo {author} {\bibfnamefont {V.}~\bibnamefont {Manga~Rao}},\
  and\ \bibinfo {author} {\bibfnamefont {S.}~\bibnamefont {Hughes}},\
  }\bibfield  {title} {\bibinfo {title} {On-chip single photon sources using
  planar photonic crystals and single quantum dots},\ }\href@noop {} {\bibfield
   {journal} {\bibinfo  {journal} {Laser Photonics Rev.}\ }\textbf {\bibinfo
  {volume} {4}},\ \bibinfo {pages} {499} (\bibinfo {year} {2010})}\BibitemShut
  {NoStop}%
\bibitem [{\citenamefont {Schwagmann}\ \emph {et~al.}(2012)\citenamefont
  {Schwagmann}, \citenamefont {Kalliakos}, \citenamefont {Ellis}, \citenamefont
  {Farrer}, \citenamefont {Griffiths}, \citenamefont {Jones}, \citenamefont
  {Ritchie},\ and\ \citenamefont {Shields}}]{schwagmann2012plane}%
  \BibitemOpen
  \bibfield  {author} {\bibinfo {author} {\bibfnamefont {A.}~\bibnamefont
  {Schwagmann}}, \bibinfo {author} {\bibfnamefont {S.}~\bibnamefont
  {Kalliakos}}, \bibinfo {author} {\bibfnamefont {D.~J.}\ \bibnamefont
  {Ellis}}, \bibinfo {author} {\bibfnamefont {I.}~\bibnamefont {Farrer}},
  \bibinfo {author} {\bibfnamefont {J.~P.}\ \bibnamefont {Griffiths}}, \bibinfo
  {author} {\bibfnamefont {G.~A.}\ \bibnamefont {Jones}}, \bibinfo {author}
  {\bibfnamefont {D.~A.}\ \bibnamefont {Ritchie}},\ and\ \bibinfo {author}
  {\bibfnamefont {A.~J.}\ \bibnamefont {Shields}},\ }\bibfield  {title}
  {\bibinfo {title} {In-plane single-photon emission from a l3 cavity coupled
  to a photonic crystal waveguide},\ }\href@noop {} {\bibfield  {journal}
  {\bibinfo  {journal} {Opt. Express}\ }\textbf {\bibinfo {volume} {20}},\
  \bibinfo {pages} {28614} (\bibinfo {year} {2012})}\BibitemShut {NoStop}%
\bibitem [{\citenamefont {Thyrrestrup}\ \emph {et~al.}(2010)\citenamefont
  {Thyrrestrup}, \citenamefont {Sapienza},\ and\ \citenamefont
  {Lodahl}}]{thyrrestrup2010extraction}%
  \BibitemOpen
  \bibfield  {author} {\bibinfo {author} {\bibfnamefont {H.}~\bibnamefont
  {Thyrrestrup}}, \bibinfo {author} {\bibfnamefont {L.}~\bibnamefont
  {Sapienza}},\ and\ \bibinfo {author} {\bibfnamefont {P.}~\bibnamefont
  {Lodahl}},\ }\bibfield  {title} {\bibinfo {title} {Extraction of the
  $\beta$-factor for single quantum dots coupled to a photonic crystal
  waveguide},\ }\href@noop {} {\bibfield  {journal} {\bibinfo  {journal} {Appl.
  Phys. Lett.}\ }\textbf {\bibinfo {volume} {96}},\ \bibinfo {pages} {231106}
  (\bibinfo {year} {2010})}\BibitemShut {NoStop}%
\bibitem [{\citenamefont {Arcari}\ \emph {et~al.}(2014)\citenamefont {Arcari},
  \citenamefont {S{\"o}llner}, \citenamefont {Javadi}, \citenamefont {Hansen},
  \citenamefont {Mahmoodian}, \citenamefont {Liu}, \citenamefont {Thyrrestrup},
  \citenamefont {Lee}, \citenamefont {Song}, \citenamefont {Stobbe},\ and\
  \citenamefont {Lodahl}}]{arcari2014near}%
  \BibitemOpen
  \bibfield  {author} {\bibinfo {author} {\bibfnamefont {M.}~\bibnamefont
  {Arcari}}, \bibinfo {author} {\bibfnamefont {I.}~\bibnamefont {S{\"o}llner}},
  \bibinfo {author} {\bibfnamefont {A.}~\bibnamefont {Javadi}}, \bibinfo
  {author} {\bibfnamefont {S.~L.}\ \bibnamefont {Hansen}}, \bibinfo {author}
  {\bibfnamefont {S.}~\bibnamefont {Mahmoodian}}, \bibinfo {author}
  {\bibfnamefont {J.}~\bibnamefont {Liu}}, \bibinfo {author} {\bibfnamefont
  {H.}~\bibnamefont {Thyrrestrup}}, \bibinfo {author} {\bibfnamefont {E.~H.}\
  \bibnamefont {Lee}}, \bibinfo {author} {\bibfnamefont {J.~D.}\ \bibnamefont
  {Song}}, \bibinfo {author} {\bibfnamefont {S.}~\bibnamefont {Stobbe}},\ and\
  \bibinfo {author} {\bibfnamefont {P.}~\bibnamefont {Lodahl}},\ }\bibfield
  {title} {\bibinfo {title} {Near-unity coupling efficiency of a quantum
  emitter to a photonic crystal waveguide},\ }\href@noop {} {\bibfield
  {journal} {\bibinfo  {journal} {Phys. Rev. Lett.}\ }\textbf {\bibinfo
  {volume} {113}},\ \bibinfo {pages} {093603} (\bibinfo {year}
  {2014})}\BibitemShut {NoStop}%
\bibitem [{\citenamefont {Vlasov}\ \emph {et~al.}(2005)\citenamefont {Vlasov},
  \citenamefont {O'boyle}, \citenamefont {Hamann},\ and\ \citenamefont
  {McNab}}]{vlasov2005active}%
  \BibitemOpen
  \bibfield  {author} {\bibinfo {author} {\bibfnamefont {Y.~A.}\ \bibnamefont
  {Vlasov}}, \bibinfo {author} {\bibfnamefont {M.}~\bibnamefont {O'boyle}},
  \bibinfo {author} {\bibfnamefont {H.~F.}\ \bibnamefont {Hamann}},\ and\
  \bibinfo {author} {\bibfnamefont {S.~J.}\ \bibnamefont {McNab}},\ }\bibfield
  {title} {\bibinfo {title} {Active control of slow light on a chip with
  photonic crystal waveguides},\ }\href@noop {} {\bibfield  {journal} {\bibinfo
   {journal} {Nature}\ }\textbf {\bibinfo {volume} {438}},\ \bibinfo {pages}
  {65} (\bibinfo {year} {2005})}\BibitemShut {NoStop}%
\bibitem [{\citenamefont {Laucht}\ \emph
  {et~al.}(2012{\natexlab{a}})\citenamefont {Laucht}, \citenamefont
  {G{\"u}nthner}, \citenamefont {P{\"u}tz}, \citenamefont {Saive},
  \citenamefont {Fr{\'e}d{\'e}rick}, \citenamefont {Hauke}, \citenamefont
  {Bichler}, \citenamefont {Amann}, \citenamefont {Holleitner}, \citenamefont
  {Kaniber},\ and\ \citenamefont {Finley}}]{laucht2012broadband}%
  \BibitemOpen
  \bibfield  {author} {\bibinfo {author} {\bibfnamefont {A.}~\bibnamefont
  {Laucht}}, \bibinfo {author} {\bibfnamefont {T.}~\bibnamefont
  {G{\"u}nthner}}, \bibinfo {author} {\bibfnamefont {S.}~\bibnamefont
  {P{\"u}tz}}, \bibinfo {author} {\bibfnamefont {R.}~\bibnamefont {Saive}},
  \bibinfo {author} {\bibfnamefont {S.}~\bibnamefont {Fr{\'e}d{\'e}rick}},
  \bibinfo {author} {\bibfnamefont {N.}~\bibnamefont {Hauke}}, \bibinfo
  {author} {\bibfnamefont {M.}~\bibnamefont {Bichler}}, \bibinfo {author}
  {\bibfnamefont {M.-C.}\ \bibnamefont {Amann}}, \bibinfo {author}
  {\bibfnamefont {A.}~\bibnamefont {Holleitner}}, \bibinfo {author}
  {\bibfnamefont {M.}~\bibnamefont {Kaniber}},\ and\ \bibinfo {author}
  {\bibfnamefont {J.~J.}\ \bibnamefont {Finley}},\ }\bibfield  {title}
  {\bibinfo {title} {Broadband purcell enhanced emission dynamics of quantum
  dots in linear photonic crystal waveguides},\ }\href@noop {} {\bibfield
  {journal} {\bibinfo  {journal} {J. Appl. Phys.}\ }\textbf {\bibinfo {volume}
  {112}},\ \bibinfo {pages} {093520} (\bibinfo {year}
  {2012}{\natexlab{a}})}\BibitemShut {NoStop}%
\bibitem [{\citenamefont {Lund-Hansen}\ \emph {et~al.}(2008)\citenamefont
  {Lund-Hansen}, \citenamefont {Stobbe}, \citenamefont {Julsgaard},
  \citenamefont {Thyrrestrup}, \citenamefont {S{\"u}nner}, \citenamefont
  {Kamp}, \citenamefont {Forchel},\ and\ \citenamefont
  {Lodahl}}]{lund2008experimental}%
  \BibitemOpen
  \bibfield  {author} {\bibinfo {author} {\bibfnamefont {T.}~\bibnamefont
  {Lund-Hansen}}, \bibinfo {author} {\bibfnamefont {S.}~\bibnamefont {Stobbe}},
  \bibinfo {author} {\bibfnamefont {B.}~\bibnamefont {Julsgaard}}, \bibinfo
  {author} {\bibfnamefont {H.}~\bibnamefont {Thyrrestrup}}, \bibinfo {author}
  {\bibfnamefont {T.}~\bibnamefont {S{\"u}nner}}, \bibinfo {author}
  {\bibfnamefont {M.}~\bibnamefont {Kamp}}, \bibinfo {author} {\bibfnamefont
  {A.}~\bibnamefont {Forchel}},\ and\ \bibinfo {author} {\bibfnamefont
  {P.}~\bibnamefont {Lodahl}},\ }\bibfield  {title} {\bibinfo {title}
  {Experimental realization of highly efficient broadband coupling of single
  quantum dots to a photonic crystal waveguide},\ }\href@noop {} {\bibfield
  {journal} {\bibinfo  {journal} {Phys. Rev. Lett.}\ }\textbf {\bibinfo
  {volume} {101}},\ \bibinfo {pages} {113903} (\bibinfo {year}
  {2008})}\BibitemShut {NoStop}%
\bibitem [{\citenamefont {Dewhurst}\ \emph {et~al.}(2010)\citenamefont
  {Dewhurst}, \citenamefont {Granados}, \citenamefont {Ellis}, \citenamefont
  {Bennett}, \citenamefont {Patel}, \citenamefont {Farrer}, \citenamefont
  {Anderson}, \citenamefont {Jones}, \citenamefont {Ritchie},\ and\
  \citenamefont {Shields}}]{dewhurst2010slow}%
  \BibitemOpen
  \bibfield  {author} {\bibinfo {author} {\bibfnamefont {S.}~\bibnamefont
  {Dewhurst}}, \bibinfo {author} {\bibfnamefont {D.}~\bibnamefont {Granados}},
  \bibinfo {author} {\bibfnamefont {D.}~\bibnamefont {Ellis}}, \bibinfo
  {author} {\bibfnamefont {A.}~\bibnamefont {Bennett}}, \bibinfo {author}
  {\bibfnamefont {R.}~\bibnamefont {Patel}}, \bibinfo {author} {\bibfnamefont
  {I.}~\bibnamefont {Farrer}}, \bibinfo {author} {\bibfnamefont
  {D.}~\bibnamefont {Anderson}}, \bibinfo {author} {\bibfnamefont
  {G.}~\bibnamefont {Jones}}, \bibinfo {author} {\bibfnamefont
  {D.}~\bibnamefont {Ritchie}},\ and\ \bibinfo {author} {\bibfnamefont
  {A.}~\bibnamefont {Shields}},\ }\bibfield  {title} {\bibinfo {title}
  {Slow-light-enhanced single quantum dot emission in a unidirectional photonic
  crystal waveguide},\ }\href@noop {} {\bibfield  {journal} {\bibinfo
  {journal} {Appl. Phys. Lett.}\ }\textbf {\bibinfo {volume} {96}},\ \bibinfo
  {pages} {031109} (\bibinfo {year} {2010})}\BibitemShut {NoStop}%
\bibitem [{\citenamefont {Ba~Hoang}\ \emph {et~al.}(2012)\citenamefont
  {Ba~Hoang}, \citenamefont {Beetz}, \citenamefont {Midolo}, \citenamefont
  {Skacel}, \citenamefont {Lermer}, \citenamefont {Kamp}, \citenamefont
  {Hofling}, \citenamefont {Balet}, \citenamefont {Chauvin},\ and\
  \citenamefont {Fiore}}]{ba2012enhanced}%
  \BibitemOpen
  \bibfield  {author} {\bibinfo {author} {\bibfnamefont {T.}~\bibnamefont
  {Ba~Hoang}}, \bibinfo {author} {\bibfnamefont {J.}~\bibnamefont {Beetz}},
  \bibinfo {author} {\bibfnamefont {L.}~\bibnamefont {Midolo}}, \bibinfo
  {author} {\bibfnamefont {M.}~\bibnamefont {Skacel}}, \bibinfo {author}
  {\bibfnamefont {M.}~\bibnamefont {Lermer}}, \bibinfo {author} {\bibfnamefont
  {M.}~\bibnamefont {Kamp}}, \bibinfo {author} {\bibfnamefont {S.}~\bibnamefont
  {Hofling}}, \bibinfo {author} {\bibfnamefont {L.}~\bibnamefont {Balet}},
  \bibinfo {author} {\bibfnamefont {N.}~\bibnamefont {Chauvin}},\ and\ \bibinfo
  {author} {\bibfnamefont {A.}~\bibnamefont {Fiore}},\ }\bibfield  {title}
  {\bibinfo {title} {Enhanced spontaneous emission from quantum dots in short
  photonic crystal waveguides},\ }\href@noop {} {\bibfield  {journal} {\bibinfo
   {journal} {Appl. Phys. Lett.}\ }\textbf {\bibinfo {volume} {100}},\ \bibinfo
  {pages} {061122} (\bibinfo {year} {2012})}\BibitemShut {NoStop}%
\bibitem [{\citenamefont {Schwagmann}\ \emph {et~al.}(2011)\citenamefont
  {Schwagmann}, \citenamefont {Kalliakos}, \citenamefont {Farrer},
  \citenamefont {Griffiths}, \citenamefont {Jones}, \citenamefont {Ritchie},\
  and\ \citenamefont {Shields}}]{schwagmann2011chip}%
  \BibitemOpen
  \bibfield  {author} {\bibinfo {author} {\bibfnamefont {A.}~\bibnamefont
  {Schwagmann}}, \bibinfo {author} {\bibfnamefont {S.}~\bibnamefont
  {Kalliakos}}, \bibinfo {author} {\bibfnamefont {I.}~\bibnamefont {Farrer}},
  \bibinfo {author} {\bibfnamefont {J.~P.}\ \bibnamefont {Griffiths}}, \bibinfo
  {author} {\bibfnamefont {G.~A.}\ \bibnamefont {Jones}}, \bibinfo {author}
  {\bibfnamefont {D.~A.}\ \bibnamefont {Ritchie}},\ and\ \bibinfo {author}
  {\bibfnamefont {A.~J.}\ \bibnamefont {Shields}},\ }\bibfield  {title}
  {\bibinfo {title} {On-chip single photon emission from an integrated
  semiconductor quantum dot into a photonic crystal waveguide},\ }\href@noop {}
  {\bibfield  {journal} {\bibinfo  {journal} {Appl. Phys. Lett.}\ }\textbf
  {\bibinfo {volume} {99}},\ \bibinfo {pages} {261108} (\bibinfo {year}
  {2011})}\BibitemShut {NoStop}%
\bibitem [{\citenamefont {Laucht}\ \emph
  {et~al.}(2012{\natexlab{b}})\citenamefont {Laucht}, \citenamefont {P\"utz},
  \citenamefont {G\"unthner}, \citenamefont {Hauke}, \citenamefont {Saive},
  \citenamefont {Fr\'ed\'erick}, \citenamefont {Bichler}, \citenamefont
  {Amann}, \citenamefont {Holleitner}, \citenamefont {Kaniber},\ and\
  \citenamefont {Finley}}]{laucht2012waveguide}%
  \BibitemOpen
  \bibfield  {author} {\bibinfo {author} {\bibfnamefont {A.}~\bibnamefont
  {Laucht}}, \bibinfo {author} {\bibfnamefont {S.}~\bibnamefont {P\"utz}},
  \bibinfo {author} {\bibfnamefont {T.}~\bibnamefont {G\"unthner}}, \bibinfo
  {author} {\bibfnamefont {N.}~\bibnamefont {Hauke}}, \bibinfo {author}
  {\bibfnamefont {R.}~\bibnamefont {Saive}}, \bibinfo {author} {\bibfnamefont
  {S.}~\bibnamefont {Fr\'ed\'erick}}, \bibinfo {author} {\bibfnamefont
  {M.}~\bibnamefont {Bichler}}, \bibinfo {author} {\bibfnamefont {M.-C.}\
  \bibnamefont {Amann}}, \bibinfo {author} {\bibfnamefont {A.~W.}\ \bibnamefont
  {Holleitner}}, \bibinfo {author} {\bibfnamefont {M.}~\bibnamefont
  {Kaniber}},\ and\ \bibinfo {author} {\bibfnamefont {J.~J.}\ \bibnamefont
  {Finley}},\ }\bibfield  {title} {\bibinfo {title} {A waveguide-coupled
  on-chip single-photon source},\ }\href@noop {} {\bibfield  {journal}
  {\bibinfo  {journal} {Phys. Rev. X}\ }\textbf {\bibinfo {volume} {2}},\
  \bibinfo {pages} {011014} (\bibinfo {year} {2012}{\natexlab{b}})}\BibitemShut
  {NoStop}%
\bibitem [{\citenamefont {Madsen}\ \emph {et~al.}(2014)\citenamefont {Madsen},
  \citenamefont {Ates}, \citenamefont {Liu}, \citenamefont {Javadi},
  \citenamefont {Albrecht}, \citenamefont {Yeo}, \citenamefont {Stobbe},\ and\
  \citenamefont {Lodahl}}]{madsen2014efficient}%
  \BibitemOpen
  \bibfield  {author} {\bibinfo {author} {\bibfnamefont {K.~H.}\ \bibnamefont
  {Madsen}}, \bibinfo {author} {\bibfnamefont {S.}~\bibnamefont {Ates}},
  \bibinfo {author} {\bibfnamefont {J.}~\bibnamefont {Liu}}, \bibinfo {author}
  {\bibfnamefont {A.}~\bibnamefont {Javadi}}, \bibinfo {author} {\bibfnamefont
  {S.}~\bibnamefont {Albrecht}}, \bibinfo {author} {\bibfnamefont
  {I.}~\bibnamefont {Yeo}}, \bibinfo {author} {\bibfnamefont {S.}~\bibnamefont
  {Stobbe}},\ and\ \bibinfo {author} {\bibfnamefont {P.}~\bibnamefont
  {Lodahl}},\ }\bibfield  {title} {\bibinfo {title} {Efficient out-coupling of
  high-purity single photons from a coherent quantum dot in a photonic-crystal
  cavity},\ }\href@noop {} {\bibfield  {journal} {\bibinfo  {journal} {Phys.
  Rev. B}\ }\textbf {\bibinfo {volume} {90}},\ \bibinfo {pages} {155303}
  (\bibinfo {year} {2014})}\BibitemShut {NoStop}%
\bibitem [{\citenamefont {Qian}\ \emph {et~al.}(2018)\citenamefont {Qian},
  \citenamefont {Wu}, \citenamefont {Song}, \citenamefont {Peng}, \citenamefont
  {Xie}, \citenamefont {Yang}, \citenamefont {Xiao}, \citenamefont {Steer},
  \citenamefont {Thayne}, \citenamefont {Tang}, \citenamefont {Zuo},
  \citenamefont {Jin}, \citenamefont {Gu},\ and\ \citenamefont
  {Xu}}]{qian2018two}%
  \BibitemOpen
  \bibfield  {author} {\bibinfo {author} {\bibfnamefont {C.}~\bibnamefont
  {Qian}}, \bibinfo {author} {\bibfnamefont {S.}~\bibnamefont {Wu}}, \bibinfo
  {author} {\bibfnamefont {F.}~\bibnamefont {Song}}, \bibinfo {author}
  {\bibfnamefont {K.}~\bibnamefont {Peng}}, \bibinfo {author} {\bibfnamefont
  {X.}~\bibnamefont {Xie}}, \bibinfo {author} {\bibfnamefont {J.}~\bibnamefont
  {Yang}}, \bibinfo {author} {\bibfnamefont {S.}~\bibnamefont {Xiao}}, \bibinfo
  {author} {\bibfnamefont {M.~J.}\ \bibnamefont {Steer}}, \bibinfo {author}
  {\bibfnamefont {I.~G.}\ \bibnamefont {Thayne}}, \bibinfo {author}
  {\bibfnamefont {C.}~\bibnamefont {Tang}}, \bibinfo {author} {\bibfnamefont
  {Z.}~\bibnamefont {Zuo}}, \bibinfo {author} {\bibfnamefont {K.}~\bibnamefont
  {Jin}}, \bibinfo {author} {\bibfnamefont {C.}~\bibnamefont {Gu}},\ and\
  \bibinfo {author} {\bibfnamefont {X.}~\bibnamefont {Xu}},\ }\bibfield
  {title} {\bibinfo {title} {Two-photon rabi splitting in a coupled system of a
  nanocavity and exciton complexes},\ }\href@noop {} {\bibfield  {journal}
  {\bibinfo  {journal} {Phys. Rev. Lett.}\ }\textbf {\bibinfo {volume} {120}},\
  \bibinfo {pages} {213901} (\bibinfo {year} {2018})}\BibitemShut {NoStop}%
\bibitem [{\citenamefont {Qian}\ \emph {et~al.}(2019)\citenamefont {Qian},
  \citenamefont {Xie}, \citenamefont {Yang}, \citenamefont {Peng},
  \citenamefont {Wu}, \citenamefont {Song}, \citenamefont {Sun}, \citenamefont
  {Dang}, \citenamefont {Yu}, \citenamefont {Steer}, \citenamefont {Thayne},
  \citenamefont {Jin}, \citenamefont {Gu},\ and\ \citenamefont
  {Xu}}]{qian2019enhanced}%
  \BibitemOpen
  \bibfield  {author} {\bibinfo {author} {\bibfnamefont {C.}~\bibnamefont
  {Qian}}, \bibinfo {author} {\bibfnamefont {X.}~\bibnamefont {Xie}}, \bibinfo
  {author} {\bibfnamefont {J.}~\bibnamefont {Yang}}, \bibinfo {author}
  {\bibfnamefont {K.}~\bibnamefont {Peng}}, \bibinfo {author} {\bibfnamefont
  {S.}~\bibnamefont {Wu}}, \bibinfo {author} {\bibfnamefont {F.}~\bibnamefont
  {Song}}, \bibinfo {author} {\bibfnamefont {S.}~\bibnamefont {Sun}}, \bibinfo
  {author} {\bibfnamefont {J.}~\bibnamefont {Dang}}, \bibinfo {author}
  {\bibfnamefont {Y.}~\bibnamefont {Yu}}, \bibinfo {author} {\bibfnamefont
  {M.~J.}\ \bibnamefont {Steer}}, \bibinfo {author} {\bibfnamefont {I.~G.}\
  \bibnamefont {Thayne}}, \bibinfo {author} {\bibfnamefont {K.}~\bibnamefont
  {Jin}}, \bibinfo {author} {\bibfnamefont {C.}~\bibnamefont {Gu}},\ and\
  \bibinfo {author} {\bibfnamefont {X.}~\bibnamefont {Xu}},\ }\bibfield
  {title} {\bibinfo {title} {Enhanced strong interaction between nanocavities
  and p-shell excitons beyond the dipole approximation},\ }\href@noop {}
  {\bibfield  {journal} {\bibinfo  {journal} {Phys. Rev. Lett.}\ }\textbf
  {\bibinfo {volume} {122}},\ \bibinfo {pages} {087401} (\bibinfo {year}
  {2019})}\BibitemShut {NoStop}%
\bibitem [{\citenamefont {Wu}\ \emph {et~al.}(2020)\citenamefont {Wu},
  \citenamefont {Peng}, \citenamefont {Xie}, \citenamefont {Yang},
  \citenamefont {Xiao}, \citenamefont {Song}, \citenamefont {Dang},
  \citenamefont {Sun}, \citenamefont {Yang}, \citenamefont {Wang},
  \citenamefont {Shi}, \citenamefont {He}, \citenamefont {Zuo},\ and\
  \citenamefont {Xu}}]{wu2020electron}%
  \BibitemOpen
  \bibfield  {author} {\bibinfo {author} {\bibfnamefont {S.}~\bibnamefont
  {Wu}}, \bibinfo {author} {\bibfnamefont {K.}~\bibnamefont {Peng}}, \bibinfo
  {author} {\bibfnamefont {X.}~\bibnamefont {Xie}}, \bibinfo {author}
  {\bibfnamefont {J.}~\bibnamefont {Yang}}, \bibinfo {author} {\bibfnamefont
  {S.}~\bibnamefont {Xiao}}, \bibinfo {author} {\bibfnamefont {F.}~\bibnamefont
  {Song}}, \bibinfo {author} {\bibfnamefont {J.}~\bibnamefont {Dang}}, \bibinfo
  {author} {\bibfnamefont {S.}~\bibnamefont {Sun}}, \bibinfo {author}
  {\bibfnamefont {L.}~\bibnamefont {Yang}}, \bibinfo {author} {\bibfnamefont
  {Y.}~\bibnamefont {Wang}}, \bibinfo {author} {\bibfnamefont {S.}~\bibnamefont
  {Shi}}, \bibinfo {author} {\bibfnamefont {J.}~\bibnamefont {He}}, \bibinfo
  {author} {\bibfnamefont {Z.}~\bibnamefont {Zuo}},\ and\ \bibinfo {author}
  {\bibfnamefont {X.}~\bibnamefont {Xu}},\ }\bibfield  {title} {\bibinfo
  {title} {Electron and hole g tensors of neutral and charged excitons in
  single quantum dots by high-resolution photocurrent spectroscopy},\
  }\href@noop {} {\bibfield  {journal} {\bibinfo  {journal} {Phys. Rev.
  Applied}\ }\textbf {\bibinfo {volume} {14}},\ \bibinfo {pages} {014049}
  (\bibinfo {year} {2020})}\BibitemShut {NoStop}%
\bibitem [{\citenamefont {Kim}\ \emph {et~al.}(2011)\citenamefont {Kim},
  \citenamefont {Shen}, \citenamefont {Sridharan}, \citenamefont {Solomon},\
  and\ \citenamefont {Waks}}]{kim2011magnetic}%
  \BibitemOpen
  \bibfield  {author} {\bibinfo {author} {\bibfnamefont {H.}~\bibnamefont
  {Kim}}, \bibinfo {author} {\bibfnamefont {T.~C.}\ \bibnamefont {Shen}},
  \bibinfo {author} {\bibfnamefont {D.}~\bibnamefont {Sridharan}}, \bibinfo
  {author} {\bibfnamefont {G.~S.}\ \bibnamefont {Solomon}},\ and\ \bibinfo
  {author} {\bibfnamefont {E.}~\bibnamefont {Waks}},\ }\bibfield  {title}
  {\bibinfo {title} {Magnetic field tuning of a quantum dot strongly coupled to
  a photonic crystal cavity},\ }\href@noop {} {\bibfield  {journal} {\bibinfo
  {journal} {Appl. Phys. Lett.}\ }\textbf {\bibinfo {volume} {98}},\ \bibinfo
  {pages} {091102} (\bibinfo {year} {2011})}\BibitemShut {NoStop}%
\bibitem [{\citenamefont {Cao}\ \emph {et~al.}(2016)\citenamefont {Cao},
  \citenamefont {Tang}, \citenamefont {Sun}, \citenamefont {Peng},
  \citenamefont {Gao}, \citenamefont {Zhao}, \citenamefont {Qian},
  \citenamefont {Sun}, \citenamefont {Ali}, \citenamefont {Shao}, \citenamefont
  {Wu}, \citenamefont {Song}, \citenamefont {Williams}, \citenamefont {Sheng},
  \citenamefont {Jin},\ and\ \citenamefont {Xu}}]{cao2016observation}%
  \BibitemOpen
  \bibfield  {author} {\bibinfo {author} {\bibfnamefont {S.}~\bibnamefont
  {Cao}}, \bibinfo {author} {\bibfnamefont {J.}~\bibnamefont {Tang}}, \bibinfo
  {author} {\bibfnamefont {Y.}~\bibnamefont {Sun}}, \bibinfo {author}
  {\bibfnamefont {K.}~\bibnamefont {Peng}}, \bibinfo {author} {\bibfnamefont
  {Y.}~\bibnamefont {Gao}}, \bibinfo {author} {\bibfnamefont {Y.}~\bibnamefont
  {Zhao}}, \bibinfo {author} {\bibfnamefont {C.}~\bibnamefont {Qian}}, \bibinfo
  {author} {\bibfnamefont {S.}~\bibnamefont {Sun}}, \bibinfo {author}
  {\bibfnamefont {H.}~\bibnamefont {Ali}}, \bibinfo {author} {\bibfnamefont
  {Y.}~\bibnamefont {Shao}}, \bibinfo {author} {\bibfnamefont {S.}~\bibnamefont
  {Wu}}, \bibinfo {author} {\bibfnamefont {F.}~\bibnamefont {Song}}, \bibinfo
  {author} {\bibfnamefont {D.~A.}\ \bibnamefont {Williams}}, \bibinfo {author}
  {\bibfnamefont {W.}~\bibnamefont {Sheng}}, \bibinfo {author} {\bibfnamefont
  {K.}~\bibnamefont {Jin}},\ and\ \bibinfo {author} {\bibfnamefont
  {X.}~\bibnamefont {Xu}},\ }\bibfield  {title} {\bibinfo {title} {Observation
  of coupling between zero-and two-dimensional semiconductor systems based on
  anomalous diamagnetic effects},\ }\href@noop {} {\bibfield  {journal}
  {\bibinfo  {journal} {Nano Res.}\ }\textbf {\bibinfo {volume} {9}},\ \bibinfo
  {pages} {306} (\bibinfo {year} {2016})}\BibitemShut {NoStop}%
\bibitem [{\citenamefont {Peng}\ \emph {et~al.}(2017)\citenamefont {Peng},
  \citenamefont {Wu}, \citenamefont {Tang}, \citenamefont {Song}, \citenamefont
  {Qian}, \citenamefont {Sun}, \citenamefont {Xiao}, \citenamefont {Wang},
  \citenamefont {Ali}, \citenamefont {Williams},\ and\ \citenamefont
  {Xu}}]{peng2017probing}%
  \BibitemOpen
  \bibfield  {author} {\bibinfo {author} {\bibfnamefont {K.}~\bibnamefont
  {Peng}}, \bibinfo {author} {\bibfnamefont {S.}~\bibnamefont {Wu}}, \bibinfo
  {author} {\bibfnamefont {J.}~\bibnamefont {Tang}}, \bibinfo {author}
  {\bibfnamefont {F.}~\bibnamefont {Song}}, \bibinfo {author} {\bibfnamefont
  {C.}~\bibnamefont {Qian}}, \bibinfo {author} {\bibfnamefont {S.}~\bibnamefont
  {Sun}}, \bibinfo {author} {\bibfnamefont {S.}~\bibnamefont {Xiao}}, \bibinfo
  {author} {\bibfnamefont {M.}~\bibnamefont {Wang}}, \bibinfo {author}
  {\bibfnamefont {H.}~\bibnamefont {Ali}}, \bibinfo {author} {\bibfnamefont
  {D.~A.}\ \bibnamefont {Williams}},\ and\ \bibinfo {author} {\bibfnamefont
  {X.}~\bibnamefont {Xu}},\ }\bibfield  {title} {\bibinfo {title} {Probing the
  dark-exciton states of a single quantum dot using photocurrent spectroscopy
  in a magnetic field},\ }\href@noop {} {\bibfield  {journal} {\bibinfo
  {journal} {Phys. Rev. Applied}\ }\textbf {\bibinfo {volume} {8}},\ \bibinfo
  {pages} {064018} (\bibinfo {year} {2017})}\BibitemShut {NoStop}%
\bibitem [{\citenamefont {Wu}\ \emph {et~al.}(2019)\citenamefont {Wu},
  \citenamefont {Peng}, \citenamefont {Battiato}, \citenamefont {Zannier},
  \citenamefont {Bertoni}, \citenamefont {Goldoni}, \citenamefont {Xie},
  \citenamefont {Yang}, \citenamefont {Xiao}, \citenamefont {Qian},
  \citenamefont {Song}, \citenamefont {Sun}, \citenamefont {Dang},
  \citenamefont {Yu}, \citenamefont {Beltram}, \citenamefont {Sorba},
  \citenamefont {Li}, \citenamefont {Li}, \citenamefont {Rossella},\ and\
  \citenamefont {Xu}}]{wu2019anisotropies}%
  \BibitemOpen
  \bibfield  {author} {\bibinfo {author} {\bibfnamefont {S.}~\bibnamefont
  {Wu}}, \bibinfo {author} {\bibfnamefont {K.}~\bibnamefont {Peng}}, \bibinfo
  {author} {\bibfnamefont {S.}~\bibnamefont {Battiato}}, \bibinfo {author}
  {\bibfnamefont {V.}~\bibnamefont {Zannier}}, \bibinfo {author} {\bibfnamefont
  {A.}~\bibnamefont {Bertoni}}, \bibinfo {author} {\bibfnamefont
  {G.}~\bibnamefont {Goldoni}}, \bibinfo {author} {\bibfnamefont
  {X.}~\bibnamefont {Xie}}, \bibinfo {author} {\bibfnamefont {J.}~\bibnamefont
  {Yang}}, \bibinfo {author} {\bibfnamefont {S.}~\bibnamefont {Xiao}}, \bibinfo
  {author} {\bibfnamefont {C.}~\bibnamefont {Qian}}, \bibinfo {author}
  {\bibfnamefont {F.}~\bibnamefont {Song}}, \bibinfo {author} {\bibfnamefont
  {S.}~\bibnamefont {Sun}}, \bibinfo {author} {\bibfnamefont {J.}~\bibnamefont
  {Dang}}, \bibinfo {author} {\bibfnamefont {Y.}~\bibnamefont {Yu}}, \bibinfo
  {author} {\bibfnamefont {F.}~\bibnamefont {Beltram}}, \bibinfo {author}
  {\bibfnamefont {L.}~\bibnamefont {Sorba}}, \bibinfo {author} {\bibfnamefont
  {A.}~\bibnamefont {Li}}, \bibinfo {author} {\bibfnamefont {B.-b.}\
  \bibnamefont {Li}}, \bibinfo {author} {\bibfnamefont {F.}~\bibnamefont
  {Rossella}},\ and\ \bibinfo {author} {\bibfnamefont {X.}~\bibnamefont {Xu}},\
  }\bibfield  {title} {\bibinfo {title} {Anisotropies of the g-factor tensor
  and diamagnetic coefficient in crystal-phase quantum dots in inp nanowires},\
  }\href@noop {} {\bibfield  {journal} {\bibinfo  {journal} {Nano Res.}\
  }\textbf {\bibinfo {volume} {12}},\ \bibinfo {pages} {2842} (\bibinfo {year}
  {2019})}\BibitemShut {NoStop}%
\bibitem [{\citenamefont {Ren}\ \emph {et~al.}(2012)\citenamefont {Ren},
  \citenamefont {Lu}, \citenamefont {Tan}, \citenamefont {Wu}, \citenamefont
  {Sun}, \citenamefont {Zhou}, \citenamefont {Xie}, \citenamefont {Sun},
  \citenamefont {Zhu}, \citenamefont {Jagadish}, \citenamefont {Shen},\ and\
  \citenamefont {Chen}}]{ren2012spin}%
  \BibitemOpen
  \bibfield  {author} {\bibinfo {author} {\bibfnamefont {Q.}~\bibnamefont
  {Ren}}, \bibinfo {author} {\bibfnamefont {J.}~\bibnamefont {Lu}}, \bibinfo
  {author} {\bibfnamefont {H.~H.}\ \bibnamefont {Tan}}, \bibinfo {author}
  {\bibfnamefont {S.}~\bibnamefont {Wu}}, \bibinfo {author} {\bibfnamefont
  {L.}~\bibnamefont {Sun}}, \bibinfo {author} {\bibfnamefont {W.}~\bibnamefont
  {Zhou}}, \bibinfo {author} {\bibfnamefont {W.}~\bibnamefont {Xie}}, \bibinfo
  {author} {\bibfnamefont {Z.}~\bibnamefont {Sun}}, \bibinfo {author}
  {\bibfnamefont {Y.}~\bibnamefont {Zhu}}, \bibinfo {author} {\bibfnamefont
  {C.}~\bibnamefont {Jagadish}}, \bibinfo {author} {\bibfnamefont {S.~C.}\
  \bibnamefont {Shen}},\ and\ \bibinfo {author} {\bibfnamefont
  {Z.}~\bibnamefont {Chen}},\ }\bibfield  {title} {\bibinfo {title}
  {Spin-resolved purcell effect in a quantum dot microcavity system},\
  }\href@noop {} {\bibfield  {journal} {\bibinfo  {journal} {Nano Lett.}\
  }\textbf {\bibinfo {volume} {12}},\ \bibinfo {pages} {3455} (\bibinfo {year}
  {2012})}\BibitemShut {NoStop}%
\bibitem [{\citenamefont {Bhattacharya}\ \emph {et~al.}(2010)\citenamefont
  {Bhattacharya}, \citenamefont {Basu}, \citenamefont {Das},\ and\
  \citenamefont {Saha}}]{bhattacharya2010quantum}%
  \BibitemOpen
  \bibfield  {author} {\bibinfo {author} {\bibfnamefont {P.}~\bibnamefont
  {Bhattacharya}}, \bibinfo {author} {\bibfnamefont {D.}~\bibnamefont {Basu}},
  \bibinfo {author} {\bibfnamefont {A.}~\bibnamefont {Das}},\ and\ \bibinfo
  {author} {\bibfnamefont {D.}~\bibnamefont {Saha}},\ }\bibfield  {title}
  {\bibinfo {title} {Quantum dot polarized light sources},\ }\href@noop {}
  {\bibfield  {journal} {\bibinfo  {journal} {Semicond Sci Technol}\ }\textbf
  {\bibinfo {volume} {26}},\ \bibinfo {pages} {014002} (\bibinfo {year}
  {2010})}\BibitemShut {NoStop}%
\bibitem [{\citenamefont {Hughes}(2004)}]{hughes2004enhanced}%
  \BibitemOpen
  \bibfield  {author} {\bibinfo {author} {\bibfnamefont {S.}~\bibnamefont
  {Hughes}},\ }\bibfield  {title} {\bibinfo {title} {Enhanced single-photon
  emission from quantum dots in photonic crystal waveguides and nanocavities},\
  }\href@noop {} {\bibfield  {journal} {\bibinfo  {journal} {Opt. Lett.}\
  }\textbf {\bibinfo {volume} {29}},\ \bibinfo {pages} {2659} (\bibinfo {year}
  {2004})}\BibitemShut {NoStop}%
\bibitem [{\citenamefont {Rao}\ and\ \citenamefont
  {Hughes}(2007)}]{rao2007single}%
  \BibitemOpen
  \bibfield  {author} {\bibinfo {author} {\bibfnamefont {V.~M.}\ \bibnamefont
  {Rao}}\ and\ \bibinfo {author} {\bibfnamefont {S.}~\bibnamefont {Hughes}},\
  }\bibfield  {title} {\bibinfo {title} {Single quantum-dot purcell factor and
  $\beta$ factor in a photonic crystal waveguide},\ }\href@noop {} {\bibfield
  {journal} {\bibinfo  {journal} {Phys. Rev. B}\ }\textbf {\bibinfo {volume}
  {75}},\ \bibinfo {pages} {205437} (\bibinfo {year} {2007})}\BibitemShut
  {NoStop}%
\bibitem [{\citenamefont {Ek}\ \emph {et~al.}(2014)\citenamefont {Ek},
  \citenamefont {Lunnemann}, \citenamefont {Chen}, \citenamefont {Semenova},
  \citenamefont {Yvind},\ and\ \citenamefont {Mork}}]{ek2014slow}%
  \BibitemOpen
  \bibfield  {author} {\bibinfo {author} {\bibfnamefont {S.}~\bibnamefont
  {Ek}}, \bibinfo {author} {\bibfnamefont {P.}~\bibnamefont {Lunnemann}},
  \bibinfo {author} {\bibfnamefont {Y.}~\bibnamefont {Chen}}, \bibinfo {author}
  {\bibfnamefont {E.}~\bibnamefont {Semenova}}, \bibinfo {author}
  {\bibfnamefont {K.}~\bibnamefont {Yvind}},\ and\ \bibinfo {author}
  {\bibfnamefont {J.}~\bibnamefont {Mork}},\ }\bibfield  {title} {\bibinfo
  {title} {Slow-light-enhanced gain in active photonic crystal waveguides},\
  }\href@noop {} {\bibfield  {journal} {\bibinfo  {journal} {Nat. Commun.}\
  }\textbf {\bibinfo {volume} {5}},\ \bibinfo {pages} {5039} (\bibinfo {year}
  {2014})}\BibitemShut {NoStop}%
\bibitem [{\citenamefont {Bayer}\ \emph {et~al.}(2002)\citenamefont {Bayer},
  \citenamefont {Ortner}, \citenamefont {Stern}, \citenamefont {Kuther},
  \citenamefont {Gorbunov}, \citenamefont {Forchel}, \citenamefont {Hawrylak},
  \citenamefont {Fafard}, \citenamefont {Hinzer}, \citenamefont {Reinecke},
  \citenamefont {Walck}, \citenamefont {Reithmaier}, \citenamefont {Klopf},\
  and\ \citenamefont {Sch\"afer}}]{bayer2002fine}%
  \BibitemOpen
  \bibfield  {author} {\bibinfo {author} {\bibfnamefont {M.}~\bibnamefont
  {Bayer}}, \bibinfo {author} {\bibfnamefont {G.}~\bibnamefont {Ortner}},
  \bibinfo {author} {\bibfnamefont {O.}~\bibnamefont {Stern}}, \bibinfo
  {author} {\bibfnamefont {A.}~\bibnamefont {Kuther}}, \bibinfo {author}
  {\bibfnamefont {A.~A.}\ \bibnamefont {Gorbunov}}, \bibinfo {author}
  {\bibfnamefont {A.}~\bibnamefont {Forchel}}, \bibinfo {author} {\bibfnamefont
  {P.}~\bibnamefont {Hawrylak}}, \bibinfo {author} {\bibfnamefont
  {S.}~\bibnamefont {Fafard}}, \bibinfo {author} {\bibfnamefont
  {K.}~\bibnamefont {Hinzer}}, \bibinfo {author} {\bibfnamefont {T.~L.}\
  \bibnamefont {Reinecke}}, \bibinfo {author} {\bibfnamefont {S.~N.}\
  \bibnamefont {Walck}}, \bibinfo {author} {\bibfnamefont {J.~P.}\ \bibnamefont
  {Reithmaier}}, \bibinfo {author} {\bibfnamefont {F.}~\bibnamefont {Klopf}},\
  and\ \bibinfo {author} {\bibfnamefont {F.}~\bibnamefont {Sch\"afer}},\
  }\bibfield  {title} {\bibinfo {title} {{Fine structure of neutral and charged
  excitons in self-assembled In(Ga)As/(Al)GaAs quantum dots}},\ }\href@noop {}
  {\bibfield  {journal} {\bibinfo  {journal} {Phys. Rev. B}\ }\textbf {\bibinfo
  {volume} {65}},\ \bibinfo {pages} {195315} (\bibinfo {year}
  {2002})}\BibitemShut {NoStop}%
\bibitem [{\citenamefont {Kuther}\ \emph {et~al.}(1998)\citenamefont {Kuther},
  \citenamefont {Bayer}, \citenamefont {Forchel}, \citenamefont {Gorbunov},
  \citenamefont {Timofeev}, \citenamefont {Sch{\"a}fer},\ and\ \citenamefont
  {Reithmaier}}]{kuther1998zeeman}%
  \BibitemOpen
  \bibfield  {author} {\bibinfo {author} {\bibfnamefont {A.}~\bibnamefont
  {Kuther}}, \bibinfo {author} {\bibfnamefont {M.}~\bibnamefont {Bayer}},
  \bibinfo {author} {\bibfnamefont {A.}~\bibnamefont {Forchel}}, \bibinfo
  {author} {\bibfnamefont {A.}~\bibnamefont {Gorbunov}}, \bibinfo {author}
  {\bibfnamefont {V.}~\bibnamefont {Timofeev}}, \bibinfo {author}
  {\bibfnamefont {F.}~\bibnamefont {Sch{\"a}fer}},\ and\ \bibinfo {author}
  {\bibfnamefont {J.}~\bibnamefont {Reithmaier}},\ }\bibfield  {title}
  {\bibinfo {title} {{Zeeman splitting of excitons and biexcitons in single
  ${\mathrm{In}}_{0.60}{\mathrm{Ga}}_{0.40}\mathrm{A}\mathrm{s}/\mathrm{G}\mathrm{a}\mathrm{A}\mathrm{s}$
  self-assembled quantum dots}},\ }\href@noop {} {\bibfield  {journal}
  {\bibinfo  {journal} {Phys. Rev. B}\ }\textbf {\bibinfo {volume} {58}},\
  \bibinfo {pages} {R7508} (\bibinfo {year} {1998})}\BibitemShut {NoStop}%
\bibitem [{\citenamefont {S{\"o}llner}\ \emph {et~al.}(2015)\citenamefont
  {S{\"o}llner}, \citenamefont {Mahmoodian}, \citenamefont {Hansen},
  \citenamefont {Midolo}, \citenamefont {Javadi}, \citenamefont
  {Kir{\v{s}}ansk{\.e}}, \citenamefont {Pregnolato}, \citenamefont {El-Ella},
  \citenamefont {Lee}, \citenamefont {Song}, \citenamefont {Stobbe},\ and\
  \citenamefont {Lodahl}}]{sollner2015deterministic}%
  \BibitemOpen
  \bibfield  {author} {\bibinfo {author} {\bibfnamefont {I.}~\bibnamefont
  {S{\"o}llner}}, \bibinfo {author} {\bibfnamefont {S.}~\bibnamefont
  {Mahmoodian}}, \bibinfo {author} {\bibfnamefont {S.~L.}\ \bibnamefont
  {Hansen}}, \bibinfo {author} {\bibfnamefont {L.}~\bibnamefont {Midolo}},
  \bibinfo {author} {\bibfnamefont {A.}~\bibnamefont {Javadi}}, \bibinfo
  {author} {\bibfnamefont {G.}~\bibnamefont {Kir{\v{s}}ansk{\.e}}}, \bibinfo
  {author} {\bibfnamefont {T.}~\bibnamefont {Pregnolato}}, \bibinfo {author}
  {\bibfnamefont {H.}~\bibnamefont {El-Ella}}, \bibinfo {author} {\bibfnamefont
  {E.~H.}\ \bibnamefont {Lee}}, \bibinfo {author} {\bibfnamefont {J.~D.}\
  \bibnamefont {Song}}, \bibinfo {author} {\bibfnamefont {S.}~\bibnamefont
  {Stobbe}},\ and\ \bibinfo {author} {\bibfnamefont {P.}~\bibnamefont
  {Lodahl}},\ }\bibfield  {title} {\bibinfo {title} {Deterministic
  photon--emitter coupling in chiral photonic circuits},\ }\href@noop {}
  {\bibfield  {journal} {\bibinfo  {journal} {Nat. Nanotechnol.}\ }\textbf
  {\bibinfo {volume} {10}},\ \bibinfo {pages} {775} (\bibinfo {year}
  {2015})}\BibitemShut {NoStop}%
\bibitem [{\citenamefont {Coles}\ \emph {et~al.}(2016)\citenamefont {Coles},
  \citenamefont {Price}, \citenamefont {Dixon}, \citenamefont {Royall},
  \citenamefont {Clarke}, \citenamefont {Kok}, \citenamefont {Skolnick},
  \citenamefont {Fox},\ and\ \citenamefont {Makhonin}}]{coles2016chirality}%
  \BibitemOpen
  \bibfield  {author} {\bibinfo {author} {\bibfnamefont {R.}~\bibnamefont
  {Coles}}, \bibinfo {author} {\bibfnamefont {D.}~\bibnamefont {Price}},
  \bibinfo {author} {\bibfnamefont {J.}~\bibnamefont {Dixon}}, \bibinfo
  {author} {\bibfnamefont {B.}~\bibnamefont {Royall}}, \bibinfo {author}
  {\bibfnamefont {E.}~\bibnamefont {Clarke}}, \bibinfo {author} {\bibfnamefont
  {P.}~\bibnamefont {Kok}}, \bibinfo {author} {\bibfnamefont {M.}~\bibnamefont
  {Skolnick}}, \bibinfo {author} {\bibfnamefont {A.}~\bibnamefont {Fox}},\ and\
  \bibinfo {author} {\bibfnamefont {M.}~\bibnamefont {Makhonin}},\ }\bibfield
  {title} {\bibinfo {title} {Chirality of nanophotonic waveguide with embedded
  quantum emitter for unidirectional spin transfer},\ }\href@noop {} {\bibfield
   {journal} {\bibinfo  {journal} {Nat. Commun.}\ }\textbf {\bibinfo {volume}
  {7}},\ \bibinfo {pages} {11183} (\bibinfo {year} {2016})}\BibitemShut
  {NoStop}%
\bibitem [{\citenamefont {Xiao}\ \emph
  {et~al.}(2021{\natexlab{a}})\citenamefont {Xiao}, \citenamefont {Wu},
  \citenamefont {Xie}, \citenamefont {Yang}, \citenamefont {Wei}, \citenamefont
  {Shi}, \citenamefont {Song}, \citenamefont {Sun}, \citenamefont {Dang},
  \citenamefont {Yang}, \citenamefont {Wang}, \citenamefont {Zuo},
  \citenamefont {Wang}, \citenamefont {Zhang},\ and\ \citenamefont
  {Xu}}]{xiao2021position}%
  \BibitemOpen
  \bibfield  {author} {\bibinfo {author} {\bibfnamefont {S.}~\bibnamefont
  {Xiao}}, \bibinfo {author} {\bibfnamefont {S.}~\bibnamefont {Wu}}, \bibinfo
  {author} {\bibfnamefont {X.}~\bibnamefont {Xie}}, \bibinfo {author}
  {\bibfnamefont {J.}~\bibnamefont {Yang}}, \bibinfo {author} {\bibfnamefont
  {W.}~\bibnamefont {Wei}}, \bibinfo {author} {\bibfnamefont {S.}~\bibnamefont
  {Shi}}, \bibinfo {author} {\bibfnamefont {F.}~\bibnamefont {Song}}, \bibinfo
  {author} {\bibfnamefont {S.}~\bibnamefont {Sun}}, \bibinfo {author}
  {\bibfnamefont {J.}~\bibnamefont {Dang}}, \bibinfo {author} {\bibfnamefont
  {L.}~\bibnamefont {Yang}}, \bibinfo {author} {\bibfnamefont {Y.}~\bibnamefont
  {Wang}}, \bibinfo {author} {\bibfnamefont {Z.}~\bibnamefont {Zuo}}, \bibinfo
  {author} {\bibfnamefont {T.}~\bibnamefont {Wang}}, \bibinfo {author}
  {\bibfnamefont {J.}~\bibnamefont {Zhang}},\ and\ \bibinfo {author}
  {\bibfnamefont {X.}~\bibnamefont {Xu}},\ }\bibfield  {title} {\bibinfo
  {title} {Position-dependent chiral coupling between single quantum dots and
  cross waveguides},\ }\href@noop {} {\bibfield  {journal} {\bibinfo  {journal}
  {Appl. Phys. Lett.}\ }\textbf {\bibinfo {volume} {118}},\ \bibinfo {pages}
  {091106} (\bibinfo {year} {2021}{\natexlab{a}})}\BibitemShut {NoStop}%
\bibitem [{\citenamefont {Xiao}\ \emph
  {et~al.}(2021{\natexlab{b}})\citenamefont {Xiao}, \citenamefont {Wu},
  \citenamefont {Xie}, \citenamefont {Yang}, \citenamefont {Wei}, \citenamefont
  {Shi}, \citenamefont {Song}, \citenamefont {Sun}, \citenamefont {Dang},
  \citenamefont {Yang}, \citenamefont {Wang}, \citenamefont {Zuo},
  \citenamefont {Wang}, \citenamefont {Zhang},\ and\ \citenamefont
  {Xu}}]{xiao2021chiral}%
  \BibitemOpen
  \bibfield  {author} {\bibinfo {author} {\bibfnamefont {S.}~\bibnamefont
  {Xiao}}, \bibinfo {author} {\bibfnamefont {S.}~\bibnamefont {Wu}}, \bibinfo
  {author} {\bibfnamefont {X.}~\bibnamefont {Xie}}, \bibinfo {author}
  {\bibfnamefont {J.}~\bibnamefont {Yang}}, \bibinfo {author} {\bibfnamefont
  {W.}~\bibnamefont {Wei}}, \bibinfo {author} {\bibfnamefont {S.}~\bibnamefont
  {Shi}}, \bibinfo {author} {\bibfnamefont {F.}~\bibnamefont {Song}}, \bibinfo
  {author} {\bibfnamefont {S.}~\bibnamefont {Sun}}, \bibinfo {author}
  {\bibfnamefont {J.}~\bibnamefont {Dang}}, \bibinfo {author} {\bibfnamefont
  {L.}~\bibnamefont {Yang}}, \bibinfo {author} {\bibfnamefont {Y.}~\bibnamefont
  {Wang}}, \bibinfo {author} {\bibfnamefont {Z.}~\bibnamefont {Zuo}}, \bibinfo
  {author} {\bibfnamefont {T.}~\bibnamefont {Wang}}, \bibinfo {author}
  {\bibfnamefont {J.}~\bibnamefont {Zhang}},\ and\ \bibinfo {author}
  {\bibfnamefont {X.}~\bibnamefont {Xu}},\ }\bibfield  {title} {\bibinfo
  {title} {Chiral photonic circuits for deterministic spin transfer},\
  }\href@noop {} {\bibfield  {journal} {\bibinfo  {journal} {Laser Photonics
  Rev.}\ }\textbf {\bibinfo {volume} {15}},\ \bibinfo {pages} {2100009}
  (\bibinfo {year} {2021}{\natexlab{b}})}\BibitemShut {NoStop}%
\bibitem [{\citenamefont {Jarlov}\ \emph {et~al.}(2015)\citenamefont {Jarlov},
  \citenamefont {Lyasota}, \citenamefont {Ferrier}, \citenamefont {Gallo},
  \citenamefont {Dwir}, \citenamefont {Rudra},\ and\ \citenamefont
  {Kapon}}]{jarlov2015exciton}%
  \BibitemOpen
  \bibfield  {author} {\bibinfo {author} {\bibfnamefont {C.}~\bibnamefont
  {Jarlov}}, \bibinfo {author} {\bibfnamefont {A.}~\bibnamefont {Lyasota}},
  \bibinfo {author} {\bibfnamefont {L.}~\bibnamefont {Ferrier}}, \bibinfo
  {author} {\bibfnamefont {P.}~\bibnamefont {Gallo}}, \bibinfo {author}
  {\bibfnamefont {B.}~\bibnamefont {Dwir}}, \bibinfo {author} {\bibfnamefont
  {A.}~\bibnamefont {Rudra}},\ and\ \bibinfo {author} {\bibfnamefont
  {E.}~\bibnamefont {Kapon}},\ }\bibfield  {title} {\bibinfo {title} {Exciton
  dynamics in a site-controlled quantum dot coupled to a photonic crystal
  cavity},\ }\href@noop {} {\bibfield  {journal} {\bibinfo  {journal} {Appl.
  Phys. Lett.}\ }\textbf {\bibinfo {volume} {107}},\ \bibinfo {pages} {191101}
  (\bibinfo {year} {2015})}\BibitemShut {NoStop}%
\bibitem [{\citenamefont {Rigal}\ \emph {et~al.}(2018)\citenamefont {Rigal},
  \citenamefont {Dwir}, \citenamefont {Rudra}, \citenamefont {Kulkova},
  \citenamefont {Lyasota},\ and\ \citenamefont {Kapon}}]{rigal2018single}%
  \BibitemOpen
  \bibfield  {author} {\bibinfo {author} {\bibfnamefont {B.}~\bibnamefont
  {Rigal}}, \bibinfo {author} {\bibfnamefont {B.}~\bibnamefont {Dwir}},
  \bibinfo {author} {\bibfnamefont {A.}~\bibnamefont {Rudra}}, \bibinfo
  {author} {\bibfnamefont {I.}~\bibnamefont {Kulkova}}, \bibinfo {author}
  {\bibfnamefont {A.}~\bibnamefont {Lyasota}},\ and\ \bibinfo {author}
  {\bibfnamefont {E.}~\bibnamefont {Kapon}},\ }\bibfield  {title} {\bibinfo
  {title} {Single photon extraction and propagation in photonic crystal
  waveguides incorporating site-controlled quantum dots},\ }\href@noop {}
  {\bibfield  {journal} {\bibinfo  {journal} {Appl. Phys. Lett.}\ }\textbf
  {\bibinfo {volume} {112}},\ \bibinfo {pages} {051105} (\bibinfo {year}
  {2018})}\BibitemShut {NoStop}%
\bibitem [{\citenamefont {Chen}\ \emph {et~al.}(2021)\citenamefont {Chen},
  \citenamefont {Fu}, \citenamefont {Gong},\ and\ \citenamefont
  {Wang}}]{chen2021quantum}%
  \BibitemOpen
  \bibfield  {author} {\bibinfo {author} {\bibfnamefont {X.}~\bibnamefont
  {Chen}}, \bibinfo {author} {\bibfnamefont {Z.}~\bibnamefont {Fu}}, \bibinfo
  {author} {\bibfnamefont {Q.}~\bibnamefont {Gong}},\ and\ \bibinfo {author}
  {\bibfnamefont {J.}~\bibnamefont {Wang}},\ }\bibfield  {title} {\bibinfo
  {title} {Quantum entanglement on photonic chips: a review},\ }\href@noop {}
  {\bibfield  {journal} {\bibinfo  {journal} {Adv.Photon.}\ }\textbf {\bibinfo
  {volume} {3}},\ \bibinfo {pages} {064002} (\bibinfo {year}
  {2021})}\BibitemShut {NoStop}%
\end{thebibliography}
\end{document}